\documentclass[iop,revtex4]{emulateapj}
\usepackage{lineno}

\newcommand{\msun}{${\cal M}_\odot$\,}

\begin{document}

\renewcommand{\topfraction}{1.0}
\renewcommand{\bottomfraction}{1.0}
\renewcommand{\textfraction}{0.0}

\title{Architecture of Hierarchical Stellar Systems and their Formation}

\author{Andrei Tokovinin}
\affil{Cerro Tololo Inter-American Observatory | NSF's NOIRLab, Casilla 603, La Serena, Chile}
\email{andrei.tokovinin@noirlab.edu}

\begin{abstract}
Accumulation  of new  data  on stellar  hierarchical  systems and  the
progress in numerical simulations of  their formation open the door to
genetic classification of these systems, where properties of a certain
group (family)  of objects are tentatively related  to their formation
mechanisms and early  evolution.  A short review of  the structure and
statistical  trends  of  known  stellar hierarchies  is  given.   Like
binaries, they can be formed by the disk and core fragmentation events
happening sequentially or simultaneously and followed by the evolution
of  masses  and orbits  driven  by  continuing  accretion of  gas  and
dynamical  interactions   between  stars.   Several   basic  formation
scenarios   are  proposed   and  associated   qualitatively   with  the
architecture of real systems,  although quantitative predictions for these
scenarios  are still  pending.  The  general trend  of  increasing orbit
alignment with decreasing  system size points to the  critical role of
the  accretion-driven   orbit  migration,  which   also  explains  the
typically  comparable   masses  of  stars belonging  to  the  same
system. The architecture of some hierarchies bears imprints of chaotic
dynamical  interactions.  Characteristic features  of each  family are
illustrated by several real systems. \\
{\it Accepted by Universe on 2021-09-18}
\end{abstract} 


\section{Introduction}
\label{sec:intro}

Formation of  stars and  planets is at  the forefront  of astronomical
research, being driven by the  need to understand our origins. Despite
tremendous progress, modern theory does not yet have enough predictive
power  to explain  statistics  of nascent  stars  from the
first  principles.   Star formation  is  a    complex and  chaotic
phenomenon involving a wide variety of inter-wined factors and physical
processes.   Stars are born  in groups,  and many  stars are  bound in
binary and  multiple systems; properties of these  systems are related
to their formation  and early evolution, and their  study advances the
general  knowledge of  star formation.   In  this review,  I focus  on
multiple  systems containing  three  or more  stars  and attempt to link  their
architecture   to  potential   formation  scenarios.  While  the
distributions of  periods, eccentricities,  and mass ratios  of binary
stars are  generally recognized as important tracers  of the star formation
mechanisms, hierarchical  systems bring additional  information encoded in
the period ratios, mutual orbit orientation, and masses of the components.

The  observational knowledge  of  multiplicity has  progressed from  a
crude description  of small and complete samples  of nearby solar-type
stars \citep{DM91,R10}  to a more detailed  assessment of multiplicity
statistics    in    different    mass    ranges    and    environments
\citep{DK13,Moe2017}.  Large  multiplicity surveys, focused  mostly on
binaries, have  revealed recently a strong  dependence of multiplicity
on mass,  and also discovered  that multiplicity statistics  depend on
the density  and temperature of  the formation environment and  on its
metallicity.  The idea that statistics of binaries (and, by extension,
stars  and  planets)  are  not  universal is  firmly  taking  root  in
astronomy and guides the research in new directions.

Theoretical studies  of star  formation also advance  rapidly.  Modern
hydrodynamical  simulations with  increasing  resolution and  improved
treatment  of  radiative  physics   and  magnetic  fields  provide  an
unprecedented and  detailed view of the formation  and early evolution
of stellar systems.   For example, \citet{Bate2019} created animations
of multiple-star  formation in a massive collapsing  cloud and pointed
out  that most  close  binaries result  from dissipative  gas-assisted
capture  of two  stars that  formed either  independently in  the same
cluster  or  in  close   proximity  (in  the  same  filament).   Other
mechanisms (disk  instability and dynamical interactions)  also play a
substantial role, and several mechanisms may combine to create a given
system.  This emerging picture is further confirmed in the simulations
by  \citet{Lee2019}  and  \citet{Kuffmeier2019}  who  studied  smaller
clouds  with a  higher resolution.   In these  simulations, companions
form sequentially, approach the main star, and migrate to closer orbits
owing to the continuing gas accretion and dynamical friction.

New knowledge is usually produced  by the joint advances of theory and
observations,  their  interaction  and  confrontation.   However,  the
complexity of  star formation, its  dependence on the  conditions, and
the  multitude of  processes  involved,  stand in  the  way of  direct
comparison   between   observations   and   theoretical   predictions.
Classification of objects helps to systematize available data and is a
necessary  step to  their  interpretation.  Its  textbook examples  in
astronomy are galaxy types and variable stars.  Classification is also
widely used in biology, relating external (observable) characteristics
of species  to their origin.  Biological  classification was developed
empirically  well before the  genetic code  became readable.   Here an
attempt  is  made  to  classify  stellar hierarchies  based  on  their
architecture and to relate  these families to the formation processes,
adopting  the   `genetic'  approach.   However,   reliable  models  of
multiple-star formation  are not yet available,  their predictions are
mostly  qualitative.  Classification  of stellar  hierarchies proposed
here  remains therefore intuitive,  reflecting the  current incomplete
understanding  of the  formation mechanisms.  I hope  that it  will be
revised and improved in the future.

Historically,   stellar  hierarchies   were  first   studied   on  the
individual, case-by-case  basis because only a few  close systems were
known, while wide visually resolved triples move too slowly on  human
time  scale.   In his  book,  \citet{Batten1973} recognized the  potential
importance  of  hierarchies   for  understanding  formation  of  close
binaries.  This  idea was  based on the  frequent occurrence  of close
binaries  in  higher-order hierarchies,  noted  by the observers.  The next  step
 was made by \citet{Fekel1981}  who assembled the first list of
35 spectroscopic binaries with tertiary companions in relatively tight
(outer periods  $<$100 yr) hierarchies  and attempted to  connect their
statistical properties with the  formation mechanisms known at that
time.   One of the  best historic  ways of  finding hierarchies  was by
detecting eclipses  in visual binaries;  a list of  80 such systems
was published by \citet{Chambliss1992}. On the other hand, several
wide physical hierarchies were documented in the catalogs of nearby
stars. 

Trying  to measure  stellar  masses  by a  combination  of visual  and
spectroscopic  orbits in  the early  1990s, I  accidentally discovered
several triples  and started  to collect data  on such  systems, being
convinced of  their astrophysical importance  by the works  of Batten,
Fekel,  and others.   The result  was the  first compilation  of known
physical hierarchical  systems covering the full range  of periods ---
the Multiple Star Catalog,  MSC \citep{MSC1}.  In parallel, systematic
spectroscopic monitoring  was initiated to  discover more hierarchies;
its results  are summarized by  \citet{TS2002}.  I extended  the 25-pc
sample  to  67\,pc  with  the  aim to  study  unbiased  statistics  of
hierarchies with solar-type  components \citep{FG67}. After completion
of the 67-pc project, I continued to update the MSC.

It  is   instructive  to  see   how  the  knowledge   of  hierarchical
multiplicity in  the solar neighborhood progressed with  time.  In the
22-pc sample  of \citet{DM91},  the estimated fraction  of hierarchies
among  all systems was  5\%.  The  more detailed  survey of  the 25-pc
sample by \citet{R10}  boosted this fraction to 13\%,  while the total
multiplicity did not change. This increase was due to the discovery of
additional  subsystems in known  binaries.  The  latest update  on the
25-pc  sample  by  \citet{Hirsch2021}   gives  the  17\%  fraction  of
hierarchies, larger  than in the earlier studies  (new subsystems were
discovered  mostly by  high-resolution imaging).   According  to these
authors, only  2/3 of  non-single stars are  pure binaries,  while the
remaining 1/3 of  the systems host three or  more stars.  Hierarchical
multiplicity  is by  no  means  rare among  solar-type  stars, and  it
increases  with mass;  almost all  massive stars  are at  least triple
\citep{Sana2017}.

I begin the review by introducing hierarchical  systems in section~\ref{sec:data},
defining  their  architecture and  the  corresponding parameters  that
serve  for  classification.   Known  statistical  trends  are  briefly
covered.   Then in section~\ref{sec:form}  the formation  mechanisms of
stars, binary stars, and  multiple systems are schematically outlined,
focusing  on   the  relation   between  formation  scenarios   and  the
architecture  of  the resulting  products.   The  proposed  classification  of
hierarchies is  presented in section~\ref{sec:class},  where each group
is  illustrated by  real systems.   Summary and  outline  of future
progress in section~\ref{sec:sum} close this review.

\section{Properties of Hierarchical Systems}
\label{sec:data}

\subsection{The Multiple Star Catalog}
\label{sec:MSC}

This study  uses data on  known hierarchical systems collected  in the
MSC; its current update is described in \citet{MSC}.  The catalog is a
compilation of random discoveries  and surveys. It is heavily affected
by  observational  selection  which  distorts  relative  frequency  of
observed  hierarchies compared  to their  intrinsic  distribution. For
example, systems  containing eclipsing binaries  are over-represented,
being  easier  to  discover.   Nevertheless,  the MSC  holds  a  large
(almost 3000)  sample  of  real multiple  systems,  allowing  to
distinguish characteristic patterns  in the multi-dimensional space of
their parameters.

Periods  and separations  of  binary and  multiple  systems span  many
orders of  magnitude, and  their other properties  such as  masses and
structure are also very  diverse. The diversity of stellar hierarchies
surpasses the diversity of exoplanetary systems.  Nowadays, exoplanets
are      successfully     classified      into      several     groups
\citep[e.g.][]{He2020}.  Here a  similar effort  to classify  the diverse
population ('zoo') of stellar hierarchies is undertaken.

The  MSC   is  regularly  updated.   Its  latest   version  is  posted
online\footnote{  \url{http://www.ctio.noirlab.edu/\~{}atokovin/stars}}
and       is       available        as  the Vizier      catalog
J/ApJS/235/6\footnote{\url{vizier.u-strasbg.fr/viz-bin/VizierR-4?-source=J/ApJS/235/6}}. Full
information  on  the multiple  systems  mentioned  in this  paper  can  be
retrieved from  the MSC. Each  system has a  unique code based  on its
J2000 coordinates, e.g. 11551+4629.  Similar codes are adopted in the
Washington Double Star Catalog, WDS \citep{WDS}, and I call them `WDS
codes' for brevity.  
Each system in the MSC has a grade; grades 3, 4, and 5 have at least
three reliably identified components,  while grades 1 and 2
(questionable and rejected systems) are not considered here. 

\subsection{Types of Hierarchy}
\label{sec:hie}

\begin{figure}
\plotone{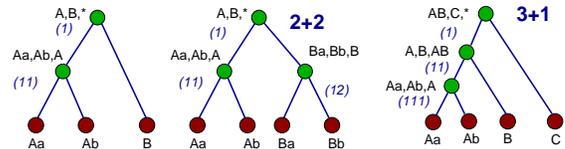}
\caption{Typical  hierarchical  structure   of  triple  and  quadruple
  systems.   Green    circles   denote   subsystems    designated   as
  (primary, secondary, parent); pink circles are stars. Numbers in brackets
  indicate hierarchical levels. 
\label{fig:levels} }
\end{figure}

A  dynamically  stable  hierarchical  system  can  be  decomposed  into  a
combination of binaries.  Its  structure is described by a binary-tree
graph,  also  called  {\em mobile  diagram.}   Figure~\ref{fig:levels}
introduces basic types  of hierarchy. A triple system  consists of the
outer (wide) pair which is at  the root of the hierarchy, and an inner
close pair.   The hierarchy can  be represented in  various equivalent
ways, e.g.  by a graph, by  numbers or levels (outer binary is level 1
and the inner  binary is level 11 or  12 if it belongs to  the main or
secondary    component,     respectively),    by    nested    brackets
\citep{Eggleton2008},  or  by reference  to  the  parent component  (a
subsystem Aa,Ab  has component  A as parent).   The first  MSC version
\citep{MSC1} coded the  hierarchy by levels, and the  current MSC uses
references to  parents, coding the  hierarchy by the  triads (primary,
secondary, parent).   The outermost subsystem  (root of the  tree) has
asterisk (*) as  parent.  Links to parent is a  flexible scheme that, on
the one hand,  defines the hierarchy and,  on the other  hand, allows easy
modifications in response to discoveries of new subsystems.

Quadruple  systems in  Figure~\ref{fig:levels} can  have  two possible
hierarchies. The 2-tier hierarchy (two close binaries orbiting each
other) is called  a 2+2  quadruple. The  3-tier hierarchy  where a
close binary  Aa,Ab has a tertiary  component B and this  triple AB is
orbited  by  another  more  distant  star  C,  is  called  a  3+1  or
`planetary' hierarchy. Indeed, it  resembles a planetary system if the
central  star   has a  substantially  larger  mass   than  its
companions.  Among solar-type stars within 67 pc,  2+2 quadruples are $\sim$4 times
more  frequent than  3+1 quadruples  \citep{FG67}. The  proportions of
single, binary, triple, quadruple  (and higher-order) systems in this
sample  are  54:29:12:5,  so  a  0.17  fraction  of  the  total
population are hierarchies. \citet{Hirsch2021} confirm this fraction for
the small but well-studied {\bf 25-pc}  sample 

\begin{figure}
\plotone{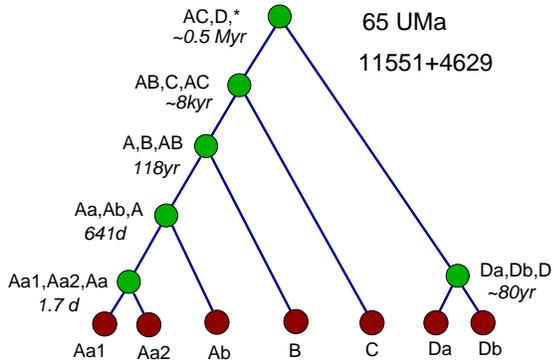}
\caption{Structure of the 5-tier hierarchical system 65 UMa
  (11551+4629, HR 4560, DN UMa). 
\label{fig:65UMa} }
\end{figure}

Complex hierarchies containing  more than four stars can  be viewed as
combinations       of        elementary       binaries,       triples,
etc. Figure~\ref{fig:65UMa}  shows the structure of  the unique 5-tier
system 65 UMa  (11551+4629); no other 5-tier hierarchies  are known so
far.  It  has an  almost  planetary-type  structure,  except that  the
outermost component D  is itself a binary. A  hierarchical system with
$N$  tiers can have  no more  than $2^N$  components if  all available
levels are filled.  This condition is fulfilled in 2+2 quadruples. The
65 UMa system has only 7 components, while the maximum possible number
of stars in a 5-tier hierarchy is  $2^5 = 32$. This situation is typical: only a
fraction of available levels are usually filled.

With  a few  exceptions, we  cannot be  sure that  all  components and
subsystems in a given hierarchy are discovered. For example, in 65~UMa
the visual components B or  C can themselves be yet undetected close
binaries or even  triples. This means that the  current description of
the  hierarchy  may  be  incomplete   and  it  will  change  with  new
discoveries. A close  spectroscopic binary Aa,Ab  with a
distant  visual companion  B looks like  a typical  triple system  with
a 2-tier  hierarchy.   However, further  study  might reveal  additional
component Ac  revolving around A at close  separation, converting this
triple into a  3+1 quadruple.  After this discovery,  the inner binary
Aa,Ab moves from  level 11 to level 111, and  the new subsystem Aab,Ac
occupies the intermediate  level 11.  

\subsection{Main Parameters of Hierarchies}
\label{sec:par}

\begin{figure*}[ht]
\plotone{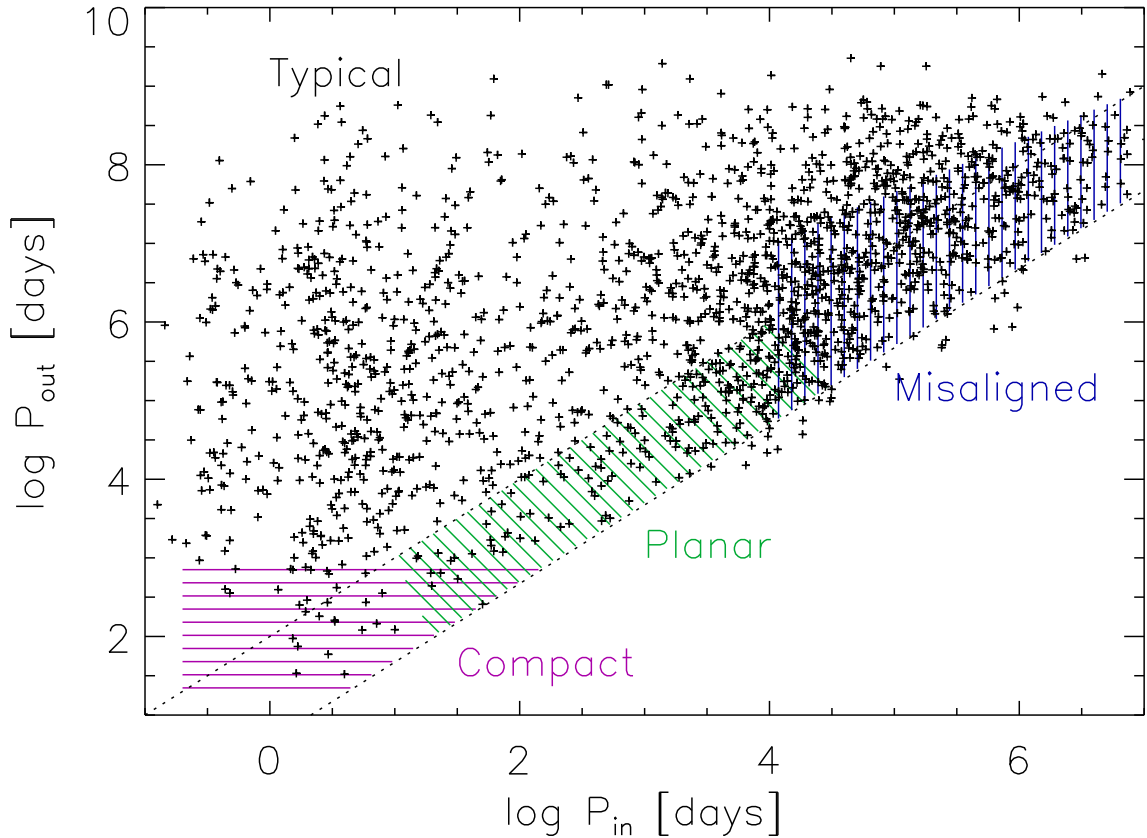}
\caption{Periods of  the  inner and  outer  binaries at  adjacent
  hierarchical levels  for known multiples  within 200 pc.  The dotted
  lines correspond  to the period ratios  of 4.7 and  100. The hatched
  areas  indicate  approximate locus  of  different classes  discussed
  further in section~3.
\label{fig:classes} }
\end{figure*}

A  hierarchical   system  can  be  decomposed   into  several  nested
binaries. Motion in each  of those binaries is approximately described
by  a Keplerian  orbit. Dynamical  interactions in  the  system change
parameters of  those instantaneous (osculating) orbits  with time, but
usually  these changes are  small and  slow.  Therefore,  the standard
orbital parameters  of a  binary --- period  $P$, semimajor  axis $a$,
eccentricity $e$,  and masses  of the components  $M_1$ and  $M_2$ ---
also  serve to  characterize  hierarchies.  A  hierarchy with  $N$
components has $N-1$ subsystems and, hence, orbits.  The masses in the
outer orbits refer to the combined masses of the inner subsystems. For
example, a  triple system containing  three equal stars will  have the
inner  mass ratio  $q_{\rm in}=M_2/M_1  =1$ and  the outer  mass ratio
$q_{\rm out}=M_3/(M_1 + M_2) = 0.5$.

Apart from the two sets of individual orbital elements, a hierarchical
triple  has additional  parameters,  namely the  period ratio  $P_{\rm
  out}/P_{\rm  in}$ and the  mutual inclination  $\Phi$ ---  the angle
between the  angular momentum  vectors of the  inner and  outer orbits
ranging  from  $0^\circ$  for  coplanar  and  co-rotating  systems  to
$180^\circ$  for  coplanar  and  counter-rotating  ones;  perpendicular
orbits  have $\Phi=90^\circ$.   Unfortunately,  mutual inclination  is
measured only  for a small subset  of known hierarchies  owing to the
observational limitations.  These  mutual parameters define the degree
and  character of dynamical  interaction between  the inner  and outer
orbits.   More   complex  hierarchies  have  even   more  such  mutual
parameters.

Figure~\ref{fig:classes} compares the inner and outer periods at adjacent
hierarchical levels.  A triple  system gives one point  in this
plot,  a quadruple  system two  points,  etc. To  reduce somewhat  the
observational selection, only 1820 systems within 200\,pc are selected
from the current MSC.  The symbol colors in the panels distinguish the systems
by the  inner mass ratio  $q_{\rm in}$ and  by the total  system mass.
The  solid  line  marks  the  limit  of  dynamical  stability  $P_{\rm
  out}/P_{\rm in} = 4.7$ (see section~\ref{sec:dyn}) and the dashed line marks $P_{\rm
  out}/P_{\rm in} = 100$. Periods  of wide binaries are estimated 
approximately  (within  a  factor  of  $\sim$3)  from  their  projected
separations using the  Kepler's third law, so some long-period triples
appear  below the  stability limit  owing to  inaccurately known  periods. This
figure illustrates the wide range  of the periods and their ratios; no
preferred ratio is apparent.


\subsection{Statistical Trends}
\label{sec:stat}

Roughly  speaking,  a triple  system  is  just  a combination  of  two
binaries, inner  and outer. The binary statistics  are relatively well
established,  at least  for solar-type  stars. Can  the  statistics of
triple  systems be  derived  from the  binary  statistics by  assuming a
random choice  of the  inner and outer  pairs from the  general binary
population  and  keeping only  dynamically  stable combinations?  This
simple approach implies that the  outer and inner pairs in a hierarchy
could be  formed independently  of each other,  and it is  called {\em
  independent  multiplicity model} (IMM).   Indeed, the  properties of
all  binaries  and  of  the  binaries  belonging  to  hierarchies  are
similar.  For example,  the  frequency of  wide  visual companions  to
spectroscopic binaries (except the closest ones) is not very different
from this frequency  for single stars.  Conversely, the  rate of close
binaries among  components of wide  visual pairs is comparable  to the
overall rate  of close  binaries in the  field. With some  caveats and
exceptions,  the  IMM roughly  matches  the  statistics of  solar-type
hierarchies \citep{FG67}, offering a crude but convenient 
mathematical recipe to compute the distribution of hierarchies over periods.  

Let $\epsilon$ be the probability that  each star is a binary. The IMM
postulates  that the frequency  of triples  should be  proportional to
$\epsilon^2$, quadruples to $\epsilon^3$, etc., because subsystems are
chosen independently  from the same parent  distribution. However, the
value  of $\epsilon$  that matches  the binary  fraction in  the solar
neighborhood  predicts  fewer  triples  than actually  observed.  This
discrepancy can  be fixed by assuming  that the field  population is a
mixture  coming  from  different  environments with  different  binary
frequencies  $\epsilon$.   Then the  relative  proportions of  single,
binary, triple, etc.  stars can  be successfully reproduced by the IMM
with  a  suitable choice  of  the  mean  $\epsilon$ and  its  variance
\citep{FG67}.  In  this paradigm,  the binary-rich constituents  of the
field contribute most binaries and hierarchies, while the binary-poor
environments  supply to  the field  mostly  single stars  and a  minor
fraction of binaries. The idea  of variable $\epsilon$ is supported by
the observational  evidences of environmental multiplicity dependence. To
give a few examples,  \citet{Deacon2020} found a measurable difference
in  the  wide-binary  frequency   between  open  clusters  and  moving
groups.  In  the   globular  cluster NGC~3201,  \citet{Kamann2020}
measured a binary frequency  of 23.1$\pm$6.1\% in the first population
of stars  and 8.2$\pm$3.5\%  in the second-generation  population that
formed later in a more dynamically active environment.

The  IMM implicitly  suggests (although,  strictly speaking,  does not
require) independent  formation of the  subsystems, therefore relative
orientation  of the  inner  and outer  orbits  in a  triple should  be
random;   there   should   be   as   many   co-rotating   triples   as
counter-rotating ones.   This assertion  can be verified  by comparing
the numbers  of apparently co- and  counter-rotating triples, assuming
their  random  orientation with  respect  to  the  observer and  equal
chances of discovering co-  and counter-rotating systems.  This method
works only  for resolved (visual)  triples in the  solar neighborhood,
where  the  sense  of  orbital   motion  in  both  subsystems  can  be
determined. This  is possible even when  only a short  fraction of the
orbit  is observed,  i.e.  for  periods of  hundreds and  thousands of
years.   An excess  of  apparently co-rotating  triples  was found  by
\citet{Worley1967}  and  later  confirmed by  \citet{Sterzik2002}  and
\citet{Tok2017}.  Therefore, the orbits of resolved visual triples are
not randomly oriented.  A more detailed study shows that the degree of
orbit  alignment depends  on  the outer  projected separation  $s_{\rm
  out}$ (in au)  and on the ratio of  periods or separations.  Compact
visual triples with  $s_{\rm out} < 50$ au  have approximately aligned
orbits;  the degree  of alignment  decreases with  separation,  and at
$s_{\rm out} > 10^3$ au the relative orbit orientation becomes random.
 

Yet another deviation  from the IMM in the  67-pc sample of solar-type
stars  is the  excess of  2+2  quadruples compared  to their  fraction
predicted by  the model \citep{FG67}.  This means  that the occurrence
of inner subsystems in both components of a wide binary is correlated.
Such  a correlation  was suspected  from the  study of  triple systems
consisting  of   a  wide  binary  and  an   inner  spectroscopic  pair
\citep{TS2002}.   Radial  velocity  (RV)  monitoring of  the  tertiary
components  in these  triples  revealed that  many  tertiaries have  a
variable  RV, i.e.  they  also contain  subsystems and  actually these
systems are 2+2 quadruples. In other words, presence of a subsystem in
one  component of  a  wide binary  enhances  the chance  of finding  a
subsystem in  the other component.  To account  for this observational
fact, the multiplicity model of the 67-pc sample postulates correlated
occurrence of the inner subsystems in both components of a wide pair.

Finally, the fact that very close (periods under 10 days) binaries are
more often  found within hierarchies (as inner  subsystems) than among
other stars obviously contradicts the IMM.  The relation between close
binaries and hierarchical systems has  been suspected a long time ago,
e.g.   by   \citet{Batten1973},  and   now  it  is   well  documented.
\citet{Tok2006}  searched  for  tertiary components  to  spectroscopic
binaries and found that their fraction anti-correlates with the binary
period: it  is close to 100\%  for binaries with periods  under 3 days
and drops progressively to 40\% at  periods of 10 days and longer; the
latter  is similar or  even lower  than the  average fraction  of wide
companions to  all stars and  proves that close binaries  {\em without
  tertiary  components}  certainly exist.  On  the  other hand, very  close
binaries prefer  to be  accompanied by other  stars, rather  than live
alone. \citet{Hwang2020} found that 14.1$\pm$1\% of contact binaries
have wide companions, while their fraction for all stars is
4.5\%. They used the Gaia catalog and made a simplifying assumption that
stars with variable fluxes are contact binaries (most of them are). 

Inverting the  argument, we  may say  that components  of wide
binaries  contain  close  pairs more frequently  than  other  stars in  the  same
population.  This  trend possibly  extends to subsystems  with periods
longer than 10 days.   \citet{Deacon2020} studied wide binaries in the
field  and  in  several  nearby  open  clusters  and  found  that  the
probability of finding a close subsystems is two times larger for the components
of wide binaries than for the average stars in the same cluster.  This
effect can be partially explained  by the variable $\epsilon$: if wide
binaries  come  preferentially from  the  binary-rich population,  the
fraction of subsystems in their  components should be also larger than
on average.  Relation betwen close and wide binaries in the Taurus
association has been studied by \citet{Joncour2017}. 

\begin{figure}
\centerline{\includegraphics[width=8.5 cm]{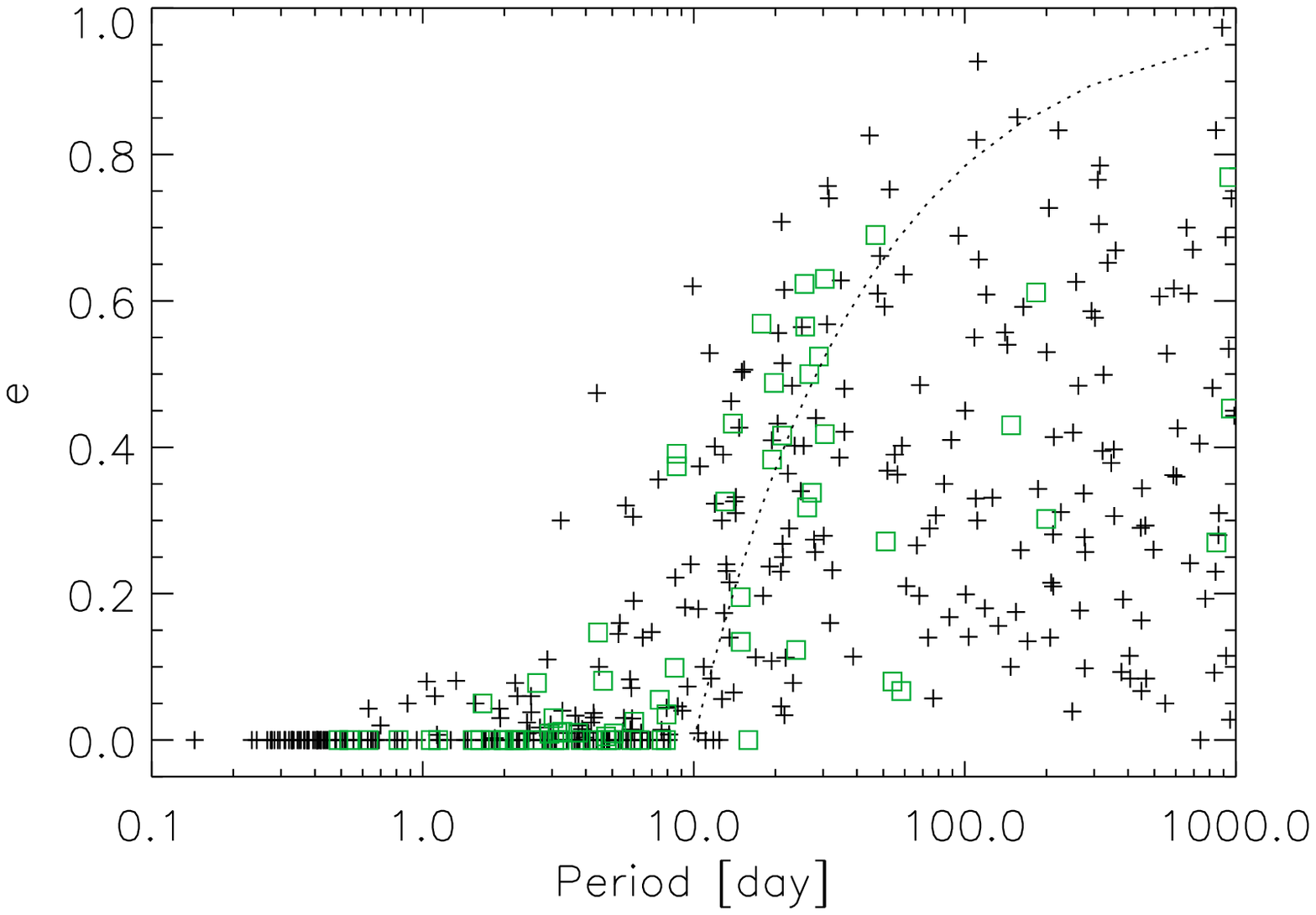} }
\centerline{\includegraphics[width=8.5 cm]{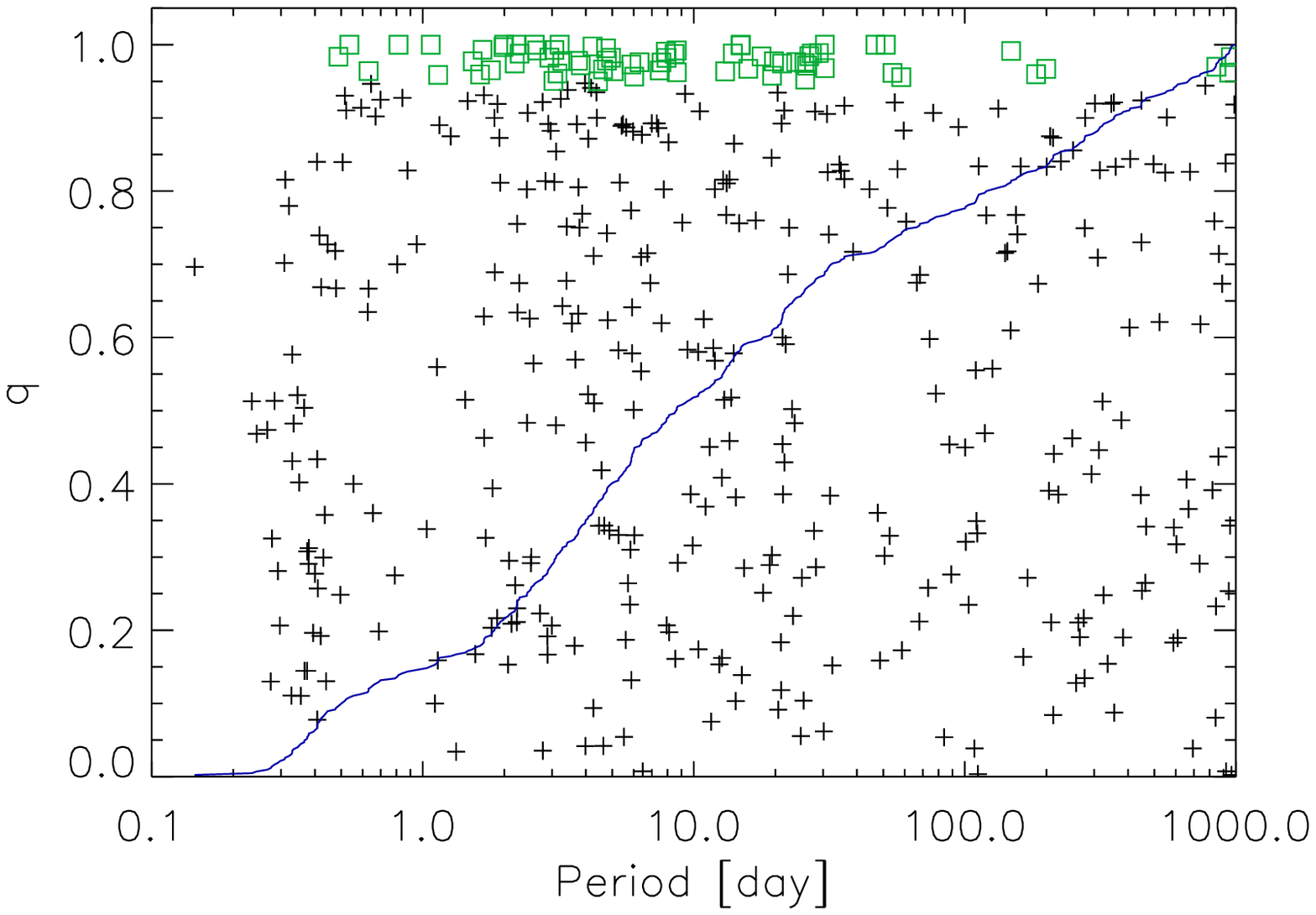}}
\caption{Properties of the inner subsystems in hierarchies. The top panel shows
  eccentricity $e$ vs. period, the bottom panel plots the inner mass
  ratio vs. period. Green squares distinguish twins with mass ratios $q>0.95$. The dotted
  line in the top panel corresponds to separation at periastron equal to 
  the radius of a circular 10-day orbit. The blue line in the lower
  panel is the cumulative period distribution. 
\label{fig:peplot} }
\end{figure}

\begin{figure*}
\plotone{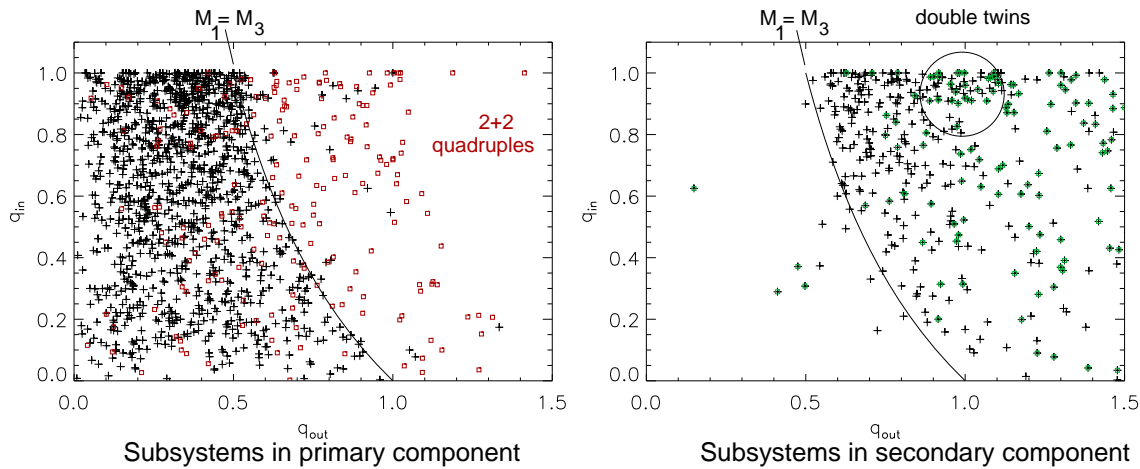}
\caption{ Relation between  inner  and outer  mass ratios  $q_{\rm
    in}=M_2/M_1$ and  $q_{\rm out}=M_3/(M_1  + M_2)$.  The  left panel
  shows  subsystems belonging  to  the primary  components, the  right
  panel --- to the  secondary components; crosses correspond to single
  outer components  and squares to outer components  that also contain
  subsystems.  The black line marks outer twins with $M_1 = M_3$.
\label{fig:qlqs} }
\end{figure*}

To take a  closer look at the {\em inner}  subsystems, I selected from
the MSC 425 inner subsystems with  primary components of less than 1.5 \msun
mass   within  200\,pc   distance  and   with  known   orbits  (mostly
spectroscopic).  Their statistics  are shown in Figure~\ref{fig:peplot}.
The  period-eccentricity plot  in  the left  panel resembles  similar
plots for  all spectroscopic binaries  \citep[e.g. figure~14 in][]{R10}.
Most  pairs with  inner periods  shorter  than 10  days have  circular
orbits owing  to the dissipative  effect of tides, although  there are
exceptions.   On the  other hand,  at  longer periods  all orbits  are
eccentric (at  $P_{\rm in} \sim  1000$ days, circular orbits  are found
again). Twin  subsystems with $q_{\rm  in} > 0.95$ (green  symbols) do
not  stand apart  from  other pairs  in  this plot.   The dotted  line
indicates the  regime where separation  at periastron is  small enough
for tidal  interaction; binaries located around this  line will evolve
to circular orbits with shorter periods \citep{Meibom2006,Moe2018}.

The  right panel  of Figure~\ref{fig:peplot}  illustrates the  lack of
correlation between period and mass ratio in the inner subsystems. The
mass ratios are measured directy for double-lined binaries and are the
minimum values for single-lined  binaries. Most twins with $q_{\rm in}
> 0.95$ have  $P_{\rm in} < 30 $  days (the same trend  exists for all
spectroscopic binaries, not just  for the inner subsystems).  The blue
line  shows the cumulative  distribution of  periods.  One  notes many
contact  systems with  $P_{\rm  in} \sim  0.3  $ days  and a  relative
deficit  of periods  between 0.5  and  2 days.   Contact binaries  are
easier to  discover by eclipses,  and their relative abundance  in the
MSC can be a selection effect.

Most  remarkably,  the  period  distribution has  no  features  around
$P_{\rm in} \sim 10 $  days. Dynamical evolution of misaligned triples
involving tides \citep[see section 3.7 and][]{Naoz2016} affects mostly
inner periods  of 10--100  days, shortening them  to $P_{\rm  in} <10$
days.  As a result, the number of subsystems in the 10--100 days range
is reduced, and they should pile  up at periods just below 10 days, as
predicted  by  \citet{Fabrycky2007}.   An  apparent  excess  of  inner
subsystems with  $P_{\rm in}  < 10$ days  noted by  \citet{TS2002} was
likely  caused   by  the   selection  (easy  discovery   of  eclipsing
subsystems), but modern, larger data do not confirm it. Absence of the
tidal signature in the distribution of the inner periods suggests that
the mechanism studied by Fabrycky  \& Tremaine could not be a dominant
factor in the formation of  close binaries, echoing the conclusions of
\citet{Moe2018}  who determine  that only  a minor  fraction  of close
binaries  coud result  from  the tidal  evolution within  hierarchies,
while most of them are products of early migration.

Our brief  overview of statistical trends would  be incomplete without
considering the mass  ratios.  Figure~\ref{fig:qlqs} compares the mass
ratios $q_{\rm in}  = M_2/M_1$ in the inner  subsystems with the outer
mass ratios  $q_{\rm out} =  M_3/(M_1+M_2)$ for the MSC  subset within
200\,pc.   Only the inner  subsystems at  the lowest  hierarchy levels
(simple  pairs of  stars without  subsystems) are  selected.  Overall,
there is no obvious correlation  between the mass ratios, although the
large number of inner subsystems with similar-mass components, $q_{\rm
  in} \sim  1$ (twins),  is notable.  Among  the 1783 inner systems  in this
plot, 483 (27\%)  belong to the secondary component  of the outer pair
(green symbols), while the remaining majority are found in the primary
component  (red  symbols).    However,  subsystems  in  the  secondary
components  are  more  difficult  to discover,  hence  their  observed
fraction is only a lower limit.

A  system where the  masses of  the tertiary  component and  the inner
primary are equal, $M_3 = M_1$, can be called {\em outer twin}, and in
such case $q_{\rm out} = 1/(1  + q_{\rm in})$.  This relation is shown
in Figure~\ref{fig:qlqs} by the black  line. When both inner and outer
mass ratios  are close to  one, the system  can be called  {\em double
  twin}.  When the tertiary star is the most massive one, $M_3 > M_1$,
the subsystem belongs to the secondary component of the wide pair, and
these  points, mostly  located to  the right  of the  black  line, are
plotted  in  green.   However,  the distinction  between  primary  and
secondary components of a wide pair does not always correlate with the
mass: it can be modified for  evolved stars, by the mass transfer, and
when the  tertiary component also  contains a subsystem (i.e.   in 2+2
quadruples). This  latter situation,  marked by squares,  explains why
some points appear  on the wrong side of the  dividing line.  A slight
concentration of  points to this line,  noted earlier \citep{Tok2008},
would  imply a  preference  of  outer twins  with  $M_3 \approx  M_1$;
however,   this   effect  does   not   appear   to  be   statistically
significant. The plot features a large number of hierarchies where all
three stars  have comparable masses  ($q_{\rm in} \approx  1$, $q_{\rm
  out} \approx 0.5$).  Remember however that the MSC data are burdened
by the observational selection, and hierarchies with comparable masses
are discovered more easily.

\begin{figure*}
\centerline{\includegraphics[width=12 cm]{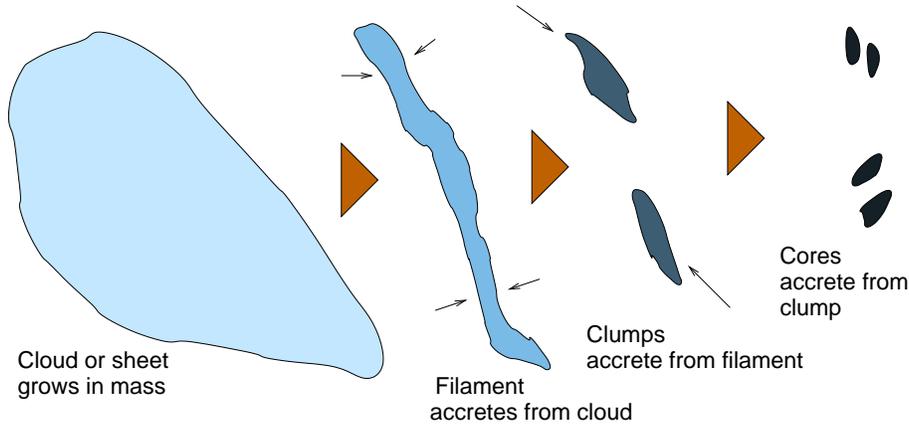}}
\caption{Illustration of  the hierarchical collapse proceeding from
  large to small scales (see the text). 
\label{fig:collapse} }
\end{figure*}

\section{Formation  of Multiple Stars}
\label{sec:form}

In this  Section,  physical  processes leading to the  formation of
binary  stars  are  briefly  reviewed  and extended  to  the  formation
scenarios of stellar hierarchies.

\subsection{Physics of Star Formation}
\label{sec:starform}

Stars  form  by  hierarchical   collapse  of  giant  molecular  clouds
\citet[see   the  review  by][]{Vasquez2019}.    It  starts  with  slow
accumulation of cold  molecular gas on large spatial  scales caused by
the Galactic spiral structure  and colliding flows in the interstellar
medium. The collapse is  highly inhomogeneous; it proceeds by creation
of two-dimensional  (sheets or slabs)  and one-dimensional (filaments)
structures.  During  collapse, the density  of the cold gas  increases and
the Jeans mass decreases, leading  to fragmentation in a cascade, from
larger  to smaller  spatial  scales (Figure~\ref{fig:collapse}).   The
fragmentation  cascade stops  when the  gas becomes  opaque  and heats
adiabatically, increasing  the Jeans mass;  the smallest scale  of the
cascade  creates stellar  embrios (protostars).   This  co-called {\it
  opacity  limit  to fragmentation}  sets  both  the  initial mass  of
the protostars  and  their minimum  separation,  on  the  order of  10  au
\citep{Larson1972}.  At  low metallicity, the gas  contains less dust,
decreasing  the opacity  and the  opacity limit  to  fragmentation and
helping to  form closer binaries.   This explains the increase  of the
close-binary  fraction at  low metallicity  found  by \citet{Moe2019}.
Interestingly,    wide   binaries    exhibit   an    opposite   trend:
\citet{Hwang2021} discovered  a reduced  fraction of wide  binaries in
low-metallicity environments  and explained  it by the  larger density
and velocity dispersion in the metal-poor star-formation regions.

Gravitational physics dictates that  collapse at small scales proceeds
much  faster than  at  large scales  \citep{Larson2007},  so when  the
smallest structures  form, the larger structures still  collapse. As a
result,  each level in  a hierarchical  collapse continues  to accrete
from the upper levels: protostars accrete from clumps, clumps from the
filament, etc. \citet{Vasquez2019} also  point out that star formation
continues on a  relatively long time scale (a few  Myr) defined by the
outermost  structure.   The  final  stellar  masses  result  from  the
accretion of gas on time scales much longer than the free-fall time of
their natal  envelopes and may be  unrelated to the  initial masses of
the  small-scale clumps  from  which the  protostars form.   Continued
accretion   on  to   newly  formed   binaries  shrinks   their  orbits
\citep{Lee2019,TokMoe2020}.    The  origins   of   the  stellar   mass
distribution  (the initial  mass function,  IMF, and  the  system mass
function,  SMF)  and its  relation  to  multiplicity  are reviewed  by
\citet{Lee2020}. The SMF can be explained as a consequence of the mass
distribution of prestellar cores (if a given fraction of the core mass
is  converted  into stars),  or  as a  result of  competitive
accretion.  \citet{Clark2021} successfully model  the observed  SMF by
assuming that turbulent fragmentation  produces low-mass seeds at some
rate and  these seeds subsequently  grow by competitive  accretion. In
their model, only 0.23 fraction of the gas mass is consumed by forming
the seeds, the rest is accreted later.

The  specific angular  momentum of  molecular gas  exceeds  by several
orders  of magnitude  the  specific angular  momentum  of a  protostar
rotating at breakup speed.  The mass growth of protostars by accretion
is  possible only  when the  angular  momentum is  extracted from  the
infalling gas.   Given that the angular momentum  transport needs some
moving mass,  accretion cannot proceed without ejecting  a fraction of
gas needed to carry away  the angular momentum excess.  The inevitable
relation between accretion and ejection  is manifested by the jets and
outflows from young stars.

\subsection{Binary Star Formation}

\begin{figure*}
\centerline{\includegraphics[width=16 cm]{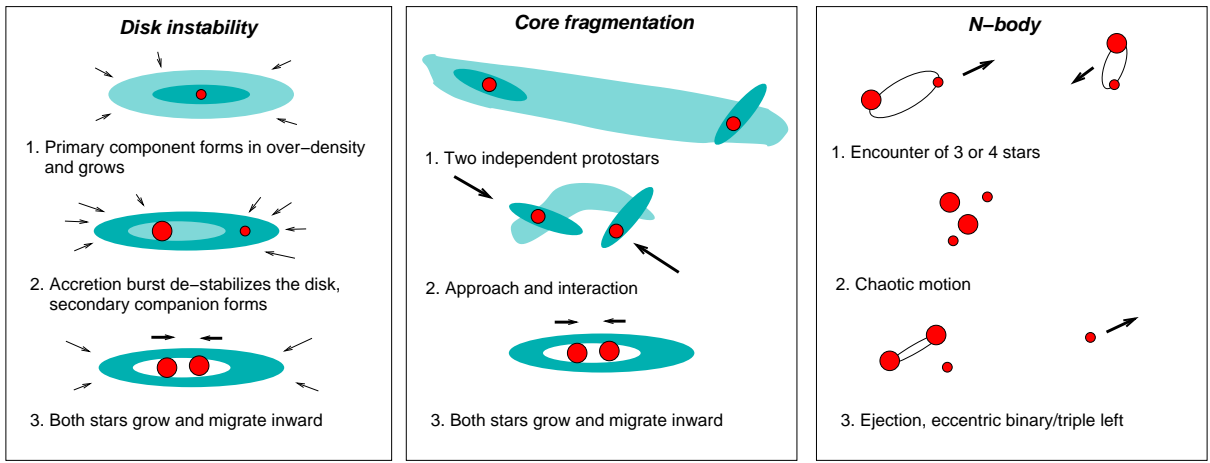}}
\caption{The two major channels of binary star formation are disk instability
  and core fragmentation. Dynamical interactions in N-body systems
  also can leave a strong imprint on the binary properties. 
\label{fig:binform} 
}
\end{figure*}

 Hierarchical systems consist of nested binaries, so mechanisms of
  binary formation and early evolution are essential for understanding
  the origin of multiples.   
Large-scale  simulation  of  collapse  by \citet{Bate2019}  show  that
binary  stars form  by  a variety  of  mechanisms acting  alone or  in
combination. Evolution of newly  formed binaries in a dense collapsing
cluster changes their parameters further by dynamical interaction with
gas and neighboring stars.  Therefore, the statistics of the resulting
binary population do  not directly relate to the  output of elementary
formation mechanisms. The two  basic channels of binary-star formation
illustrated in Figure~\ref{fig:binform}  are disk instability and core
fragmentation.

{\it  Disk instability}  (DI).  As  already noted,  star  formation is
impossible  without shedding  the  excessive angular  momentum of  the
infalling gas.  If  the angular momentum transport is  slower than its
influx, the accreted  gas accumulates in a disk  around the protostar.
A massive disk can become  unstable to fragmentation, forming one or
several companions  around the nascent  central protostar \citep[][and
  references therein]{Kratter2010}. The opacity limit to fragmentation
dictates that  the initial  separations of binaries  formed by  the DI
mechanism are larger than $\sim$10 au.

Disks sufficiently massive to become unstable are more likely to exist
around massive stars  \citep{Kratter2010}. However, accretion is known
to be highly  variable (episodic). A burst of  accretion increases the
disk mass temporarily and may lead to the disk fragmentation even when
the average  disk is too  small to fragment.   The opacity of  the gas
plays  an  important  role  here,  and  the  DI  is  enhanced  at  low
metallicity \citep{Moe2019}.

A binary formed by DI has initially a very small mass ratio. Continued
accretion on to such a binary increases masses of both components, but
the  secondary  component  grows  faster  and the  mass  ratio  always
increases.  At  the same time,  the orbit shrinks  through interaction
with  the circumbinary  disk  \citep{Heath2020}. The  accretion-driven
binary migration is a  complex process depending on several  factors, such as
orientation  of the binary orbit  relative to the angular  momentum of
the  infalling  gas,  size  of  the  inner  circumstellar  disks,  gas
temperature, etc.  Statistical modeling of accreting binaries based on
simplified prescriptions  can reproduce the overall  properties of the
close-binary population \citep{TokMoe2020}.  The final mass ratio of a
DI-formed  binary   correlates  with  the  time   of  the  companion's
formation: binaries that  formed early tend to have  larger mass ratios
and develop  a sub-population of  twins with nearly  identical masses.
Conversely, companions  formed by the end of  the mass-assembly period
have  little chance  to grow.   The  DI mechanism  explains the  large
fraction  of close  binaries found  among massive  stars by  the large
amount of gas needed to form such stars, favoring binary formation and
migration.   Furthermore, strong  migration can  lead to  mergers of
binary components,  which is more likely for  massive binaries; early binary
mergers can produce the most massive stars.

{\it  Core  fragmentation}  (CF)   is  probably  the  dominant  binary
formation mechanism,  being a  direct consequence of  the hierarchical
collapse. The last  stage of the collapse is set  by the opacity limit
to  fragmentation  ($\sim$10  au)   which  also  defines  the  minimum
separation  of   binaries  that  can  be  formed   by  CF.   Numerical
simulations  of   an  isolated  cloud  collapse   driven  by  internal
turbulence  show how  several  protostars usually  are  born from  the
individual over-densities  \citep{Offner2010,Lee2019}.  Given that the
gravitational  collapse creates  two- and  one-dimensional structures,
the  protostars can  form  in linear  configurations along  filaments.
Pairs  of   neighboring  protostars  may   end  up  in   bound  binary
systems.  Over-densities are  usually produced  by gas  compression in
converging flows, where  the kinetic energy of the  colliding flows is
mutually canceled.   Relative velocities of  the clumps in  a filament
are  therefore smaller  than  the typical  large-scale gas  velocities
\citep{Kuffmeier2019}.  Two  protostars born in the  same filament can
be gravitationally bound from the outset if their relative velocity is
less than the escape velocity \citep{Tok2017}.

Simulations   of  \citet{Bate2019}   show   that even originally   unbound
protostars  in a  cluster  can  end  up in  a  binary, while  the
excessive kinetic  energy of their  encounter is dissipated by  the gas.
Dissipative capture is the dominant mechanism of binary formation in 
these simulations.  It should  also work in isolated collapsing filaments.
Two neighboring  protostars approach each  other while falling  to the
common center of mass.  At  the same time, each protostar continues to
attract and accrete gas.   When the two protostars become sufficiently
close to each  other, their envelopes interact, the  kinetic energy is
dissipated, and they  form a gravitationally bound pair  with a common
envelope, even if their  initial relative velocity exceeded the escape
velocity.  The initial  size of such binaries should  be comparable to
the size of their disks or envelopes, on the order of a few hundred au
or less.

A  binary pair  formed by  CF continues  to evolve  and migrate  as it
accretes more gas \citep{Lee2019}. This is similar to the evolution of
a DI-formed binary, except that  the initial spatial scales can be larger
and the motions  of the gas relative to the binary  are more likely to
be chaotic, as happens in nascent clusters. The end products of the CF
and  DI mechanisms can  be very  similar.  In  a dense cluster,  a competing
mechanism  of binary  evolution  are dynamical  interactions with  other
stars or binaries. They modify the orbits, can disrupt the binary, and
often involve exchanges of the components \citep{Bate2019}.

Binaries help  to form  stars by storing  the angular momentum  of the
infalling  gas in  their orbits.   \citet{Sterzik2003} noted  that the
specific angular  momentum  of  cores  is  comparable to  the  typical  angular
momentum  of binaries.   According to  their logic,  low-mass binaries
originate  from small  cores  and have  correspondingly small  orbits,
matching the observed trend in the binary separations vs. mass. If the
size of binary orbits is  indeed determined mostly by the core angular
momentum, we can talk  about rotationally-driven core fragmentation in
the  spirit of  the early  collapse simulations  by \citet{Larson1972}.
Fragmentation  of  disk-like  or  bar-like  rotating  structures  also
happens  in the  modern  cluster simulations  by \citet{Bate2019},  as
evidenced  by movies provided  in the  supplementary material  to that
paper.

The most  efficient way to store  the excessive angular  momentum in a
binary orbit  is by  forming two fragments  of comparable masses  on a
near-circular  orbit, because  this configuration  corresponds  to the
maximum specific (i.e. per unit mass) orbital angular momentum.  As if
by coincidence, these characteristics  are typical for the lowest-mass
binaries   \citep{Dupuy2011}.   On  the   other  hand,   formation  of
companions to  low-mass stars  by DI is  unlikely \citep{Kratter2010}.
Therefore,  fragmentation  of isolated  low-mass  cores  might be  the
dominant  mechanism of  forming low-mass  binaries.  \citet{Rohde2021}
simulated a  large number of fragmenting  cores of 1  \msun total mass
and  found that  the resulting  binaries have  a strong  preference to
equal  masses,  including  many   twins.   The  typical  mass  of  the
components, $\sim$0.3 \msun, corresponds to M-type dwarfs (only a half
of  the  initial  core  mass  was converted  into  stars).   In  these
simulations, episodic  outflows are shown to  be important, decreasing
the average  final masses and  yielding a more  realistic multiplicity
statistics. On the  other hand, variations of the  core radius, virial
parameter, and turbulence spectrum  had little effect on the resulting
masses and multiplicity. These simulations also produced a substantial
number of triples, some of which were dynamically unstable.

The existence of close (spectroscopic) low-mass binaries and the large
fraction  of the low-mass  twins suggest  that the  CF-formed binaries
also  migrate.  The  toy  model of  accretion-driven binary  migration
proposed by \citet{TokMoe2020} is  equally applicable to the CF-formed
binaries.   When a  CF-formed  binary accretes  substantial mass,  the
resulting close  pair has the  same properties as the  DI-formed close
binary, erasing the signature  of its initial formation mechanism. One
may  even  wonder  whether  the  distinction between  the  DI  and  CF
mechanisms of binary formation is as meaningful as it appears at the first
sight.

\subsection{Formation Scenarios of Hierarchical Systems}
\label{sec:multform}

Hierarchical systems  with three or  more components can be  formed by
the  combination  of   elementary mechanisms  outlined  above  and
further modified  by the N-body interactions.  The  sequence of events
producing a  hierarchy is called {\em  formation scenario}.  Potential
scenarios     are     outlined      below     and     summarized     in
Table~\ref{tab:scenarios}. The  proposed scenarios are  hypothetical and
qualitative. Examples of these scenarios  can be found in the numerical
simulations of star formation. In  all cases, gas plays a critical role
in the formation of hierarchies.

\begin{table*}
\caption{Formation Scenarios
\label{tab:scenarios} }
\center
\begin{tabular}{l l l c c l}
\hline
Scenario & Predictions & Products \\
\hline
Sequential  disk instability  & Aligned orbits with moderate            & 1. Compact planar triples \\
 ~~~(DI+DI)                   & eccentricity, $q_{\rm in} \ge q_{\rm out}$, & 2. Double-twin triples \\
                    & moderate perod ratios,                            & 3. Quadruples of 3+1 type \\
                    & no  2+2 quadruples.                               & \\
Sequential core fragmentation & Non-coplanar, eccentric orbits.         & 1. Typical triples \\ 
~~~(CF+CF, DI+CF)             & Wide range of mass ratios.              & 2. Quadruples 2+2  \\        
Late disk instability         & Small  $q_{\rm in}$, misaligned           & Castor-type quadruples \\
~~~(CF+DI2)                   & inner subsystems.                       & \\
Cloud collisions              & Wide 2+2 quadruples,            & $\epsilon$~Lyr-type quadruples \\
                              & comparable masses              &    \\
Dynamical interactions        & Eccentric and                  & Misaligned triples \\ 
                              & misaligned orbits, small             & \\
                              & period ratio                            &  \\
\hline
\end{tabular}
\end {table*}

\begin{figure}
\includegraphics[width=8.5 cm]{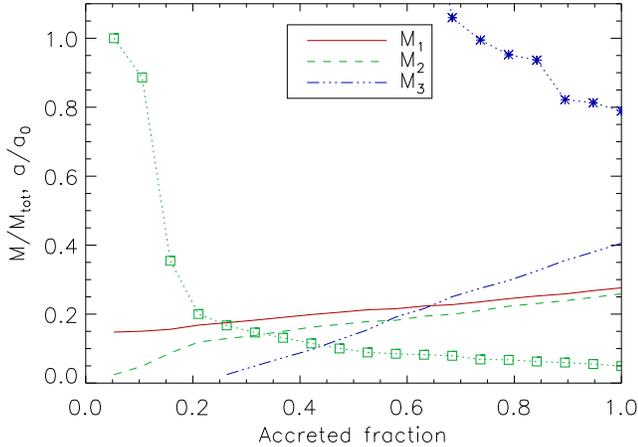}
\caption{Formation of a triple system by sequential disk fragmentation
  and  migration according  to the  model of  \citet{TokMoe2020}.  The
  lines show the  fraction of mass in each star  vs. fraction of total
  accreted mass.  The squares and  asterisks show the inner  and outer
  semimajor axes,  respectively, in  relative units. The  first binary
  (green) forms early, at 0.05 accreted fraction, and migrates from 30
  au to  0.05 au,  reaching the  final inner masses  of 0.80  and 0.75
  \msun  (a twin).   The tertiary  component  (blue) forms  at 178  au
  separation (outside the plot limit)  when 0.25 fraction of the  mass was
  already accreted and migrates to 23 au,  reaching the mass of 1.18 \msun and
  becoming  the  primary component  of  the  triple.   In most  cases,
  however, the  tertiary forms  later and does  not outgrow  the inner
  two stars.
\label{fig:trip} 
}
\end{figure}

\subsection{Sequential disk instability (DI+DI)}  

While a close  binary formed by DI migrates  inward, another companion
can be  formed by DI  on its periphery \citep{TokMoe2020}.   The DI+DI
process can create compact triple systems and 3+1 quadruples.  In this
scenario,  hierarchies  are assembled  inside-out,  starting from  the
innermost  and closest  binary. The  accreted mass  is
preferentially retained  by the least massive  outer component because
it moves  on a  wider orbit around  the center  of mass, but  when the
components  grow   and  become   comparable,  the  accreted   mass  is
distributed evenly, leading to the  formation of twins with $q \approx
1$.    This   consideration   applies   to  both   inner   and   outer
subsystems. The DI+DI scenario makes a strong prediction regarding the
outer mass ratio:  it cannot exceed one.  This  limit is attained when
most mass is accreted after formation of the triple. In such case,
the resulting system is a {\em double twin} where both inner and outer
mass ratios  are close  to one  and the masses  of the  components are
distributed  approximately in  the 1:1:2  proportion. 

The  outer, most  massive component  in  a double  twin outshines  the
fainter inner binary and reduces the chance of its discovery.  Several
double   twins   were   discovered   only  recently   using   advanced
observational  techniques  \citep{Tok2018b}.  Double twins  are  still
relatively rare compared to other known hierarchies (note the points near (1,1)
in Figure~\ref{fig:qlqs}).  A case  where the outer companion grows to
become the most massive star in  a triple system (but not yet a double
twin)  is  illustrated  in  Figure~\ref{fig:trip}.   However,  in  the
majority of  simulated DI+DI triples  the outer companion  forms later
and remains the  least massive star of the  system.  The distributions
of  periods  and  mass  ratios  of simulated  triples  are  plotted  in
\citet{TokMoe2020}. \citet{Tobin2016} found a triple protostar where a
low-mass tertiary component is apparently forming in the circumbinary disk. 

Growth of  the outer companion can be stopped  or slowed down when
yet another, fourth star forms at  a larger separation, thus converting a
nascent triple into a  3+1 planetary-type quadruple.  Indeed, in the known
planetary-type quadruples  the tertiary and  quaternary components are
usually less massive than the central pair.

The  DI+DI   scenario  implies   a  planar  or   quasi-planar  orbital
architecture if the angular momentum  of the accreted gas maintains 
approximately  constant  direction during  the  mass assembly  period.
Otherwise, accretion  of misaligned  gas on to  a triple  or quadruple
modifies the orientation  of  its  orbits and  destroys  the  coplanarity,
partially or totally.  Yet another consequence of the accretion is the
inward  migration of  the orbits.   A newly  formed  low-mass tertiary
component  overtakes the  accretion flow,  previously directed  to the
inner binary, and migrates  faster. Later, if the masses become comparable,
the gas is distributed more evenly and the inner binary also migrates.
The inner and outer orbits, evolving jointly to closer separations, 
interact dynamically and can be trapped in a mean motion resonance (MMR), at
least  temporarily \citep{Tremaine2020}.  There  is some  analogy with
the  migration in  multi-planet systems  that also  produces sometimes
resonant or quasi-resonant architectures.  On the other hand, if the tertiary
component migrates faster than the inner binary, the system can become
dynamically unstable and will be disrupted, ejecting some stars.

\subsection{Core fragmentation  and capture (CF+CF, DI+CF).}

A close  binary can form by either core  fragmentation or  disk instability
if it accretes  enough gas and migrates.  Then  another component formed
at a  large distance can be  captured, producing a triple  system in a
DI+CF or CF+CF  scenario, as found  in the  simulations by  \citet{Lee2019},
\citet{Kuffmeier2019}, and \citet{Rohde2021}.   A  variant of  this  scenario is  independent
formation of  two pairs by either  DI or CF mechanisms and their 
encounter.  If all gas is exhausted or removed by the time of the
encounter, the  dynamical interaction  between two binaries  can leave
behind a  triple system and an  ejected star.  If, on  the other hand,
the interaction  is moderated by  gas, both binaries can  survive with
modified orbits \citep{Ryu2017}.

Elongated  filamentary structures  produced by  gravitational collapse
tend to accumulate mass near their ends by means of the so-called edge
collapse. \citet{Yuan2020} found that  the structure of the 25-pc long
filament S228  matches the edge collapse  predictions.  This mechanism
can create two adjacent star clusters or groups of stars of comparable
masses. Edge  collapse is a consequence of  the gravitational focusing
that attracts  gas to  the opposite ends  of elongated  structures and
drives the  longitudinal gas motion.  This mechanism  should also work
at smaller scales, producing wide binaries or multiples from elongated
cores.   In a  series of  papers, \citet{Bonnell1993}  explored binary
formation from elongated cores and compared predictions of their early
simulations  with  observations.  \citet{Sadavoy2017} found  that  the
orientation of wide pairs  of protostars in elongated cores correlates
with the core axis at separations exceeding $\sim$500 au, while closer
pairs  are   oriented  randomly  with  respect  to   the  core.  These
observations  hint  that  protostars  likely  form near  the  ends  of
elongated structures and fall to the center of mass.

In  the CF+CF  scenario, the  inner  and outer  subsystems are  formed
independently of  each other, therefore  their orbits can  be oriented
randomly and  the component's  masses can span  a wide  range. Typical
triple  systems  with  wide  outer orbits  qualitatively  match  these
predictions.   However, in  the hierarchical  collapse  each structure
accretes gas  on a  time scale  longer than its  free-fall time,  so a
hierarchy  formed  by  the  CF+CF  scenario continues  to  evolve  and
migrate. As  in the DI+DI  scenario, accretion of a  substantial mass
shrinks the  orbits and  equalizes the masses.   Moreover, dissipative
interaction  with gas  reduces  the eccentricities  and increases  the
orbit alignment.   An initially wide and misaligned  2+2 quadruple can
end up as a compact and  aligned system of four equal-mass stars. This
seems  to  be the  only  way  to explain  the  origin  of compact  2+2
quadruples with outer separations much less than the opacity limit for
fragmentation.

One can envision a cascade of fragmentations where most of the angular
momentum remains in  the mutual orbit of two  fragments which collapse
and further fragment,  forming a quadruple system of  2+2 hierarchy in
the  outside-in fragmentation  sequence  \citep{Bodenheimer1978}.  For
efficient  storage of  the angular  momentum in  the  fragments, their
masses should  be similar.  Although such 2+2  architecture is typical
of  many  multiple systems  consisting  of  four  stars of  comparable
masses,  most  modern  hydrodynamical  simulations do  not  favor  the
cascade hierarchical fragmentation scenario, featuring instead chaotic
gas  motions and  sequential (rather  than simultaneous)  formation of
protostars.  In these simulations, hydrodynamic and magnetic transport
of the angular momentum is  sufficient to enable accretion and buildup
of stellar  masses.  However, recent simulations of  a solar-mass core
fragmentation by \citet{Rohde2021}  yield some triples with comparable
(small) masses and aligned  orbits resembling the expected products of
a fragmentation cascade.

\subsection{Collisions}

Mutual interactions between nascent  protostars play an important role
in dense clusters, leading  to captures, exchanges, disruptions, etc.,
as  demonstrated  in  the  simulations of  \citet{Bate2019}.  However,
mutual interactions  also can take  place in a  low-density star-forming
regions because they are  highly structured and the local density in the clumps 
can  be high even  when the  average density  is low.  However, unlike
cluster stars, members of small groups interact with each other for a short
time, only during their first encounter \citep{Delgado-Donate2003}.

Interactions  between gravitating point masses without gas (`dry') are
discussed   in  the   following  subsection.    In   the  star-forming
environments,  gas  is  dynamically  important.  Collisions  at  large
spatial  scales that compress the  interstellar  medium  and trigger  star
formation \citep{Vasquez2019} can be relevant at smaller scales
as well. A collision between  two starless cores creates a shock front
and can lead to the collapse of each core into a binary, forming a 2+2
quadruple system in a single  event.  This mechanism has been proposed
and simulated by \citet{Whitworth2001}.

An  encounter  of two accreting  protostars  can  produce a  multiple
system in several  ways. The encounter perturbs the  gas envelopes and
produces a  burst of accretion  on to the  stars that can lead  to the
formation  of inner  subsystems by  the DI  mechanism  (enhancement of
accretion by  a fly-by or  collision is well  known in the  context of
galaxy  evolution).  At  the  same  time, the  kinetic  energy of  the
relative motion is dissipated and the two initially unbound protostars
become a  wide bound pair.  The result  can be a 2+2  quadruple, as in
the   collision   between   starless   cores,   or   just   a   triple
system. Simultaneous presence of subsystems in both components of wide
binaries, noted  above, suggests that  the subsystems originated  in a
common event, which  could be a collision of  protostars surrounded by
the  gas envelopes or  a collision  of starless  cores.  While  in the
DI+DI  or CF+CF  scenarios the  hierarchies are  built  from inside-out
(inner pairs form first), the collision-formed hierarchies are born
in one event.

Along with collisions between  cores or protostars, one might envision
a  collision between  a relatively  wide binary  and a  gas  cloud. An
episode of enhanced accretion  promotes buildup of circumstellar disks
and can  create  subsystems  around   each  binary  component  by  the  DI
mechanism, converting the binary into a 2+2 quadruple.  In this CF+DI2
(late   disk  instability)  scenario,   the  hierarchy   is  assembled
outside-in,  starting from  the outermost  pair and  adding  the inner
subsystems  later.  With a  limited gas  supply, the  inner companions
will not accrete much and the inner mass ratios will remain small. The
angular momentum of  the gas in the accretion  burst is not necesarily
aligned  with the outer  binary, so  the two  inner subsystems  can be
mutually  aligned  but  not  coplanar  with  the  outer  orbit.   This
hypothetic  scenario  matches  properties of  Castor-type  quadruples,
where the inner subsystems have small mass ratios, while the two outer
primaries have comparable masses. A variant of this scenario where the
gas cloud  colliding with  the wide binary  also contains  a protostar
could produce a sextuple system.

\subsection{Dynamical Interactions}
\label{sec:dyn}

To  the first  approximation, stellar  hierarchies can  be  modeled by
gravitationally  interacting point masses  (an N-body  system).  Their
dynamics is well  understood, despite the lack of  analytic theory for
systems of three or more gravitating points. The motion can be studied
by means  of the perturbation  approach that represents the  inner and
outer  systems  by  Keplerian  two-body orbits  with  slowly  changing
(osculating) elements.  This approximation works well for systems with
a strong hierarchy, i.e. with a large ratio of periods or separations.
Otherwise, the motion can be explored by direct numerical integration.
A   recent   review   of   triple-star  dynamics   is   published   by
\citet{Docobo2021}, while the book by \citet{Valtonen2006} is
recommended for a deeper study of the three-body problem. 

When the  separations between three masses are  comparable, the triple
system  is  non-hierarchical   and  dynamically  unstable;  a  regular
quasi-Keplerian motion is replaced  by the chaotic `interplay'.  Close
triple approaches  or `scrambles' lead to  ejections of one  star on a
wide  orbit, and during  ejections the  system temporarily  appears as
hierarchical,      until      the      next     interplay      episode
\citep{Anosova1986,Manwadkar2020}.  Eventually,  one star (usually the
least massive  one) is  ejected from the  system, while  the remaining
binary  shrinks.  Assuming  that the  interplay erases  memory  of the
initial   conditions,  \citet{Stone2019}   derived   the  eccentricity
distribution  of  binaries  that  remain  after decay of unstable triples. 

The distinction  between non-hierarchical (unstable)  and hierarchical
(stable)  triple systems has  been studied  by many  authors.  Several
criteria of  dynamical stability of triple systems  based on numerical
simulations have  been proposed in  the literature.  For  example, the
criterion of \citet{Mardling2001} for coplanar orbits can be recast as
\begin{equation}
P_{\rm out}/P_{\rm in} > 4.7 (1 - e_{\rm out})^{-1.8}  (1 + e_{\rm out})^{0.6}  (1 + q_{\rm  out})^{0.1} ,
\label{eq:Mardling}
\end{equation}
where $e_{\rm out}$ is the eccentricity of the outer orbit and $q_{\rm
  out}$ is  the ratio  of the distant-companion  mass to  the combined
mass of the inner binary. The limiting period ratio of 4.7 corresponds
to  the solid  line  in Figure~\ref{fig:classes}.   There exist  multiple
systems with  period ratios  close to this  limit (e.g.   LHS~10170 and
$\zeta$~Aqr presented below),  hence with strong dynamical interaction
between the inner and outer orbits.

Dynamically  unstable multiple  systems  with comparable  separations,
so-called {\em  trapezia}, are short-lived, making  their discovery in
mature stellar populations unlikely.  The best chance to find trapezia
is  by studying  the pre-main  sequence (PMS)  stars;  indeed, several
interesting young  trapezia candidates  are known.  Dynamical  decay of
young unstable  multiples leads to the ejection  of components (single
stars  or tight  binaries) with  a  high speed,  on the  order of  the
orbital speed at the moment  of close interaction.  Runaway stars that
move  away from  young  star-formation regions  witness the  dynamical
decay of young multiples, while  their velocities provide an order of
magnitude  estimate of  the size  of those  systems at  the  moment of
disruption.  Existence  of hierarchies with moderate  period ratios is
an indirect evidence that some unstable young systems  have decayed,
while the  stable and marginally-stable ones survived.

Triple  systems  that experienced a   chaotic  dynamical  evolution  bear
characteristic  imprints of  this  process: their  orbits are  usually
misaligned and have large  eccentricities, while the period ratios are
moderate (not  too far from  the stability limit). These  features are
inferred from the  simulations of cluster decay \citep{Sterzik2002} and
from the  numerical scattering experiments  \citep{Antognini2016}. The
eccentricity distribution is thermal,  $f(e)=2e$, or even steeper than
thermal \citep{Stone2019}. 

Stars  typically form  in  small groups  with comparable  separations.
These  small-N  clusters  should  evolve dynamically,  ejecting  some
members. When such cluster is  immersed in a massive gas cocoon, the
ejected star can be pulled back  by the additional gravity of the gas
and  return, instead  of being  lost.   \citet{Reipurth2012} suggested
such  ejections as  a mechanism  for  forming very  wide binaries  and
called it {\em unfolding}.  In  this scenario, each wide binary should
contain an  inner close pair  (unless that pair migrated  strongly and
merged),  explaining the observed  correlation between  wide multiples
and triples.  The  orbit of the unfolded tertiary  is necessarily very
eccentric because  its angular momentum  is comparable to  the angular
momentum  of the  initial unstable  triple which,  according  to these
authors, can have  a size of a few hundred au.   If the wide companion
repeatedly  returns to  the  central  gas cloud,  its  orbit might  be
circularized  and the  triple will  no longer  be wide.   The scenario
proposed by  Reipurth \& Mikkola  is elegant, but some  predictions of
the  unfolding mechanism  do not  match observations.   In populations
younger than  a few  Myr, wide binaries  are already  abundant, whereas
unfolding   postulates    their   delayed   formation.     The   outer
eccentricities of  wide triples are  not always large,  signaling that
unfolding  is unlikely to  be the  dominant channel  of wide  binary (or
triple)    formation;   there    are   other,    less    exotic   ways
\citep{Tok2017b}.

When the  distance between binary components  becomes small, comparable  to the
stellar radii,  the stars  can no longer  be treated as  point masses.
Tidal   forces  cause   precession  of   the  inner   orbit   and  its
circularization.    A  relatively  wide   triple  with   large  mutual
inclination,  $39^\circ < \Phi  < 141^\circ$,  experiences Lidov-Kozai (LK)
cycles that  modulate both $\Phi$  and $e_{\rm in}$  \citep[see][for a
  review]{Naoz2016}. The increased  inner eccentricity may `switch on'
tidal  interaction in  the  inner  orbit near  its  periastron.  As  a
result,  the LK cycles  break  up   and  the  inner  orbit  shrinks  and
circularizes.   This  mechanism of  close-binary  formation is  called
Kozai cycles with  tidal friction, KCTF \citep{KCTF,Eggleton2006}.  It
can be viewed  as a kind of migration where the  angular momentum of the
inner subsystem is transferred to the orbit of the tertiary companion.
Formation of close  binaries by the KCTF mechanism matches   their preference to
be  inner  subsystems in  multiples.   This  is  particularly true  at
periods shorter than a few  days: such binaries cannot form very early
because the  radii of PMS  stars are large  and the binary  would have
merged early on.  The strong tendency of close  binaries to be members
of hierarchical  systems \citep{Tok2006} finds its  explanation in the
KCTF  mechanism,  at  least partially.  \citet{Fabrycky2007}  explored
statistics  of  close  binaries  formed  by  this  channel, although
\citet{Moe2018}  argue  that it  cannot  be  the dominant mechanism of
close-binary formation. 

Dynamical interactions in mature hierarchical systems are
astrophysically important and can produce some exotic objects
\citep{Hamers2020,Toonen2020}. However, they are outside the scope of
this review. 

\section{Families of Hierarchical Systems}
\label{sec:class}

In  this Section,  I  explore the  relation  between   hypothetical
formation scenarios of hierarchies  and the properties of real systems,
trying  to develop  their `genetic'  classification. The  knowledge of
formation  processes  is  still  qualitative,  their  predictions  are
tentative,  and   a given  system  can  be  associated with  several
formation  scenarios.  Hopefully, this  attempt will  stimulate further
theoretical and observational work in this area.

\begin{figure*}
\centerline{\includegraphics[width=16 cm]{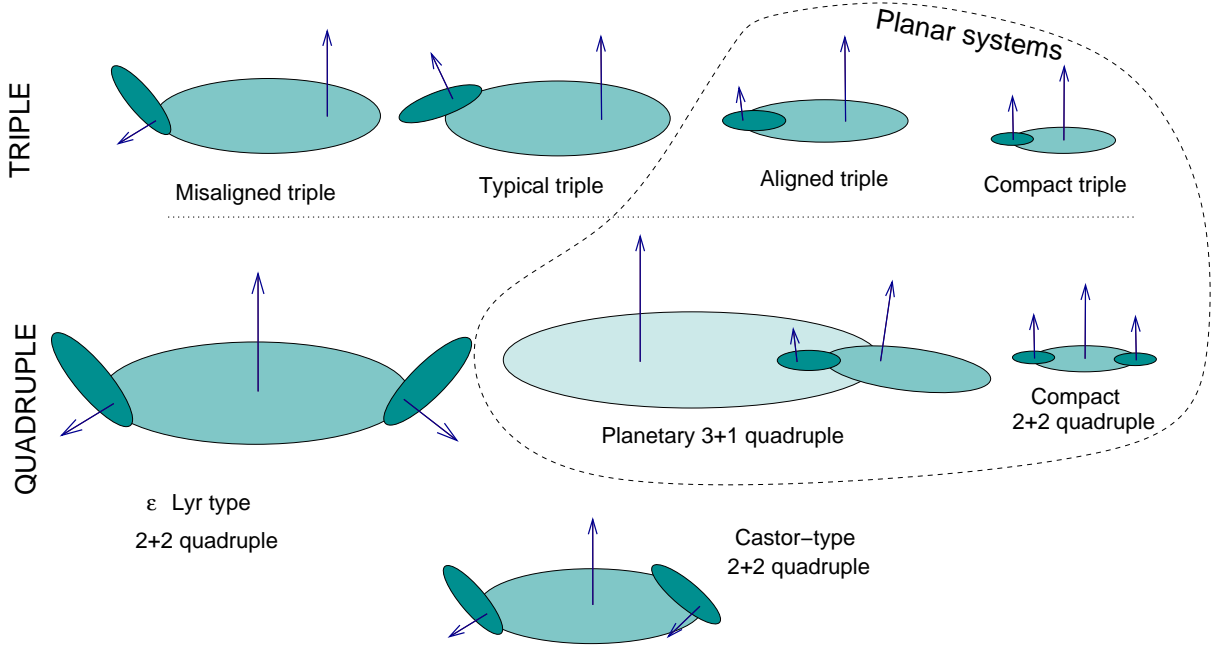}}
\caption{Cartoon illustrating the classification of hierarchical triples
  and quadruples. Ellipses depict the orbits, arrows indicate the orbital
  angular momentum vectors. Typical separations decrease from left to
  right, while the degree of orbit alignment increases and the orbits
  become less eccentric.
\label{fig:families} }
\end{figure*}

Figure~\ref{fig:classes}  reproduces  the  $P_{\rm in},  P_{\rm  out}$
diagram and marks the location of groups discussed below. However, the
classification  is  based  not  only  on the  periods,  but  on  other
parameters (mutual orbit  orientation, eccentricities, mass ratios) as
well, so  different groups  overlap in this  diagram.  The  upper left
corner is occupied by the  hierarchies with short inner and long outer
periods, $P_{\rm  out}/ P_{\rm in}  > 100$.  These hierarchies  can be
called {\em  typical}.  Owing to the large  period ratios, interaction
between their inner and  outer orbits is negligible.  Hierarchies with
smaller  period  ratios  are   more  interesting,  and  the  following
discussion is focused mostly on  these systems.  To help visualize the
forthcoming   text,   Figure~\ref{fig:families}  shows   schematically
various families of triple and quadruple systems.

\begin{table*}
\caption{Selected Multiple Systems
\label{tab:mult} }
\begin{tabular}{l l l c c l}
\hline
Group & WDS & Names & $N$ & H & Periods \\
\hline
Planar         & 00247$-$2653  & LHS 1070, GJ 2005   & 3  & 3   & 99yr(18.2yr)  \\
Large-N        & 01158$-$6853  & $\kappa$~Tuc, GJ 55.3   & 5 & 3+2       & 270kyr(1.2kyr(22yr);85yr) \\
Compact        & 01379$-$8259  & HIP 7601       & 3 & 3       & 1.75yr(19.4d) \\
Misaligned     & 02460$-$0457  & HD 17251, ADS 2111       & 3 & 3 & 1kyr(38.5yr) \\
Compact        & 03009$-$3427  & TIC 209409435  & 3 & 3       & 121.9d(5.7d) \\
Misaligned     & 04007$+$1229 & $\lambda$~Tau, HR 1239    & 3  & 3 & 33d(4d) \\
Castor         & 04141$-$3155  & TIC 168789839  & 6 & (2+2)+2 & 1.6kyr(3.9yr(1.6d;1.3d);8.2d)  \\
Compact        & 04191$+$0054  & TIC 454140642            & 4 & 2+2 & 432d(13.6d;10.4d) \\ 
Castor?        & 04357$+$1010  & HR~1458, 88 Tau & 6 & (2+2)+2 & 80kyr(18yr(3.6d;7.9d);3.7yr) \\ 
Misaligned     & 05387$-$0236  & $\sigma$~Ori, ADS 4241   & 3+ & 3 & 157yr(143d) \\
Castor         & 07346$+$3153  & Castor, $\alpha$~Gem   & 7 & (2+2)+3 & 15kyr(460yr(9.2d;2.9d);55yr(0.8d)) \\
Planar         & 08270$+$2157  & HIP 41431, GJ 307      & 4 & 3+1     & 3.9yr(59d(2.9d)) \\
Large-N        & 08316$+$1924  & HIP 41824, GJ 2069, CU~Cnc    & 5 & 3+2       & 2kyr(66yr(2.8d);23yr) \\
$\epsilon$ Lyr & 08571$-$2951  & HIP 43947      & 4 & 2+2 & 14kyr(106yr;100yr) \\
Planar         & 10370$-$0850  & HD 91962       & 4 & 3+1     & 205yr(8.8yr(170d)) \\
Compact        & 11029$+$3025  & VW LMi, HD 95660         & 4 & 2+2     & 355d(7.9d;0.5d) \\
Large-N        & 11551$+$4629  & 65 UMa, ADS 8347         & 7 & \ldots    & 500kyr(8kyr(118yr(641d(1.7d)));80yr) \\
Large-N        & 16044$-$1122  & $\xi$~Sco, ADS 9909       & 5 & 3+2       & 285kyr(1.5kyr(46yr);4.4kyr) \\
Large-N        & 16555$-$0820  & HIP 81817, GJ 644, Wolf 630   & 5 & 3+1+1     & 50kyr(9kyr(1.7yr(3d))) \\
Planar         & 16057$-$3252  & HIP 78842      & 4 & 3+1     & 4.4kyr(133yr(10.5yr)) \\
Compact        & 16073$-$2204  & HD 144548               & 3+ & 3      & 33.9d(1.63d) \\ 
Compact        & 17521$-$2920  & OGLE BLG-ECL-145467 & 4  & 2+2 & 4.2yr(4.9d;3.3d) \\    
Compact        & 18278$+$2442  & V994 Her, ADS 11373       & 6 & (2+2)+2 & 1.4kyr(2.9yr(2.1d;1.4d);2d) \\
$\epsilon$ Lyr & 18443$+$3940  & $\epsilon$ Lyr & 4 & 2+2 & 450kyr(1804yr;724yr) \\
$\epsilon$ Lyr & 18455$+$0530  & FIN 332, ADS 11640        & 4 & 2+2 & 3kyr(40yr;28yr) \\
Compact        & 19499$+$4107  & KIC 5897826, KOI-126    & 3 & 3       & 33.9d(1.8d) \\
Miasligned     & 20396$+$0458  & HIP 101955, GJ 795       & 3 & 3 & 38.7yr(2.5yr) \\
Misaligned     & 22288$-$0001  & $\zeta$~Aqr, ADS 15971    & 3 & 3 & 540yr(26yr) \\
Compact        & 22366$-$0034  & HIP 111598     & 3 & 3       & 271d(5.9d) \\
\hline
\end{tabular}
\end{table*}

Table~\ref{tab:mult}  gives   the  main  characteristics   of  several
selected   hierarchical    systems   discussed   in    the   following
subsections. The  information is extracted  from the MSC.  The systems
are identifed by the WDS codes and common names. Then the total number
of known components  $N$ and the hierarchy type H  are given. The last
column gives the orbital periods in the bracket notation, like $P_{\rm
  out}( P_{\rm in1}; P_{\rm in2}) $ for a 2+2 quadruple.

\subsection{Planar Systems}
\label{sec:plan}

There  is  a  general  trend  of  increasing  orbital  alignment  with
decreasing  outer separation.  Visual  triples with  outer separations
less  than $\sim$50  au tend  to  have aligned  orbits: their  average
mutual  inclination   $\Phi$  is  about   $20^\circ$  \citep{Tok2017}.
Moreover, in  the corotating (hence likely aligned)  systems the inner
eccentricities are,  on average, smaller than  in the counter-rotating
systems.  Planar architecture and moderate eccentricities suggest that
these systems were  possibly formed by the DI+DI  scenario.  The inner
binary forms first,  grows in mass and evolves  to closer separations,
while  the  outer  companion(s)  form  later.   Further  accretion  of
substantial mass  by such a triple  should produce a  double twin with
both inner  and outer mass ratios  close to one.   Several such visual
triples  were  identified  in  \citet{Tok2018b}  and  called  `dancing
twins'. Moderate period ratios in these systems imply strong dynamical
interaction between the inner and outer orbits.

One of the most remarkable dancing twins is the low-mass triple system
LHS~1070 (00247$-$2653). The main star A has a mass of 0.12 \msun, and
the inner subsystem B,C is composed  of two stars of 0.075 \msun each,
on   the   borderline  between   hydrogen-burning   stars  and   brown
dwarfs. Both inner and outer mass  ratios are close to one.  The inner
period is  18.2 yr and the  outer period (still preliminary)  is 99 yr
according to  \citet{Xia2019}. Both inner and outer  orbits are nearly
circular and coplanar ($\Phi =  1.7^\circ$).  The period ratio of 5.44
is  close to  the  limit of  dynamical  stability; however,  numerical
integration  performed   by  Xia  et  al.  confirms   stability  on  a
$\sim$1~Gyr  time scale.   \citet{Tok2018b} found  earlier  the period
ratio of  4.5, suggestive  of a  9:2 MMR.  A  longer time  coverage is
needed to  better constrain  the outer period  and thus to  confirm or
refute  the MMR  hypothesis.   Interaction between  the orbits  causes
measurable deviations from  the Keplerian motion in the  inner pair of
LHS~1070. The structure of  LHS~1070 resembles low-mass triples formed
by   fragmentation   of  isolated   cores   in   the  simulations   of
\citet{Rohde2021}.

Double  twins  are  difficult  to discover  because  the  bright
  tertiary outshines  the faint inner pair.  This  means  than only a
handful  of them  are studied  so far.  Newly discovered  double twins
still  lack the  coverage  of  the outer  orbits.   Most known  planar
triples  with  outer separations  below  $\sim$50  au  are not  double
twins,nhowever. Their mutual inclinations are measured when both inner
and  outer  orbits  are  resolved,  either  directly  or  through  the
astrometric  wobble. A special  case where  the mutual  inclination is
deduced from dynamical modeling  of compact triples is discussed
below in section~\ref{sec:compact}.

As noted above, the growth of the tertiary companion is stopped when a
more  distant fourth companion  forms and  overtakes the  accreted gas
flow. In such case, the result is a quasi-planar `planetary' hierarchy
of  3+1 type.  An example  of such  system is  HD~91962 (10370$-$0850)
studied by  \citet{Tok2015}.  It  is called planetary  because three
small companions  of K,  M spectral types  revolve around  the central
most  massive star  of G0V  type and  also because  their  orbits have
moderate eccentricities (0.30, 0.125, 0.135, counting outside-in). The
mutual inclination between the  outer and middle orbits is $11^\circ$,
and  the less  certain mutual  inclination between  the  innermost and
middle  orbits,  determined  later   by  the  astrometric  wobble,  is
$32^\circ   \pm   12^\circ$.    The   mutual  inclinations   and   the
eccentricities in  HD~91962 are larger  than in the solar  system, but
smaller than in  typical multiples. The ratio of  the middle and inner
periods, $18.97 \pm  0.06$, is close to an integer  number, but a 1:19
MMR is very weak, hence unlikely.

The 3+1  quadruple HIP~78842 (16057$-$3252) has the  inner triple with
periods  of 10.5 and  131 yr  (period ratio  12.6), again  with nearly
coplanar  and  circular  orbits  and  with  a  mutual  inclination  of
$12^\circ$  \citep{Tok2018b}. The  inner  pair Ba,Bb  is  a twin  with
masses of  0.75 \msun each, but  the mass of  the main star A  is only
0.96 \msun,  so this is  not a double  twin.  The reason might  be the
existence  of  a fourth  companion  C  at  9.3$''$ separation  (the
estimated period  of AB,C is $\sim$4  kyr) with a mass  of 0.64 \msun.
The parameters  of the outer orbit  cannot be determined  owing to its
long  period.  However,  the  small eccentricity  of the  intermediate
131-yr orbit, $e= 0.03$, implies the absence  of the LK cycles, so the
outer mutual inclination should  be less than $39^\circ$.  Compared to
HD~91962,  this planetary-type quadruple  is much  wider, and  all its
subsystems are resolved.

On the  other hand, a compact  3+1 quadruple system  composed of three
very similar  K7V dwarfs and  a lower-mass outer  companion, HIP~41431
(08270+2157),  was discovered  and  studied by  \citet{Borkovits2019}.
The orbital  periods are  2.9 days,  59 days, and  3.9 yr.  The mutual
inclination  between the  two inner  orbits is  2.2$^\circ$, and  the outer
orbit  is inclined  by $\sim$21$^\circ$.   The inner  pair  is eclipsing.
Accurate eclipse  timing reveals strong  dynamical interaction between
all three  orbits.  The intermediate 59-day orbit  precesses under the
influence of  the outer  4-yr companion.  The  system is  not resolved
visually,  the   mutual  inclinations  are   determined  by  dynamical
modeling.

Summarizing,  the family  of  planar hierarchies  is distinguished  by
their  approximately  aligned  orbits  ($\Phi <  40^\circ$),  moderate
period  ratios   $P_{\rm  out}/  P_{\rm   in}  <  50$,   and  moderate
eccentricities  ($e<0.5$).   The outer  separations  are also  modest,
typically  $<50$ au.  These features  suggest formation  by  the DI+DI
scenario. Resonances predicted by this  scenario are not yet proven to
exist in the real systems. Although \citet{Quirrenbach2019} discovered
that the giant star $\nu$~Oph is  orbited by two brown dwarfs in a
MMR  (periods 503  and  1385 days,  ratio 1:6),  the  small mass  ratios are  more
typical  for planetary  systems.   The planar  family includes  double
twins and  3+1 planetary quadruples,  but the majority of  its members
are more common  triples. A subset of compact  planar hierarchies with
short outer periods is discussed in the following section.

\subsection{Compact Hierarchies}
\label{sec:compact}

Looking at Figure~\ref{fig:classes}, we note a relatively small number of
points with $P_{\rm out} < 10^3$ days in the lower left corner of the
diagram. Rareness of such hierarchies is even more evident in the
volume-limited sample of solar-type stars \citep{FG67}. I call these
systems {\em compact}; they  belong to the planar family, as the
orbits are normally aligned. An example of a compact planar system is
HIP~41431 (08270+2157) presented above \citep{Borkovits2019}.

Close  tertiary  companions  to  spectroscopic  binaries  are  readily
detectable by   variation of the systemic  velocity.  Despite this,
such cases are rare owing to the intrinsically low fraction of compact
hierarchies.   Among  several  known  examples  of  compact  (but  not
eclipsing)   spectroscopic   triples,   I   can   mention   HIP   7601
(01379$-$8259,  periods   1.75  yr  and   19.4  days)  and   HIP  15988
(2366$-$0034, periods  271 and  5.9 days).  

Tertiary  companions to close binaries can be discovered by the
astrometric  acceleration  they produce.  One  of  the most  sensitive
probes is the so-called proper motion anomaly (PMA) --- the difference
between  the  long-term  proper  motion  deduced  from  the  positions
measured by  the Gaia and  Hipparcos space missions, and  the accurate
short-term  motion measured  by  Gaia \citep{Brandt2018,Kervella2019}.
Examination  of the  PMA  of close  binaries  in the  67-pc sample  of
solar-type stars revealed  a few dozen new triples  that were added to
the MSC.  However, the  periods of the astrometric tertiary companions
are unknown, and only a fraction of those triples could be compact.

The best way  to detect and study compact hierarchies  is based on the
analysis  of eclipses  in the  inner subsystems.   A large  number of such
hierarchies were discovered by eclipse timing variations in the sample
of Kepler eclipsing binaries by \citet{Borkovits2016}. They are indeed
close  to coplanarity,  with  a few  exceptions.   The Kepler  compact
hierarchies  are mostly  beyond 200\,pc  from  the Sun,  and for  this
reason they are not  plotted in Figure~\ref{fig:classes}.  This technique
has led  to the discovery of ultra-planar compact hierarchies.
For example, \citet{Borkovits2020} found a triple system TIC~209409435
with periods  of 121.9  and 5.7 days  where the mutual  inclination is
constrained  to  be  $\Phi   <  0.25^\circ$.   Several  other  compact
ultra-planar systems  were discovered recently by  this method. One
of the  shortest outer  periods, 33.9 days,  is found for  KIC 5897826
(KOI-126, 19499+4107); the inner period  is 1.8 days \citep{Borkovits2016}. The
young triply  eclipsing system HD  144548 (16073$-$2204) in  the Uppper
Scorpius association  has very similar  periods of 33.9 and  1.63 days
\citep{Alonso2015}. \citet{Borkovits2020} give a list of 17 compact
triples contaning eclipsing binaries; their outer periods range from
33 to 1100 days.

The most compact among known  2+2 quadruples is VW~LMi (11029+3035, HD
95660) with an outer period of only 355 days \citep{Pribulla2020}. The
inner  periods are 7.93  and 0.48  days. All  four stars  have similar
masses of  $\sim$1 \msun and are  tightly packed in an  outer orbit of
only 1.6 au size. The  coplanarity of orbits in this unresolved system
is inferred from the  dynamical analysis.  The doubly eclipsing system
V994 Her  (18278+2442) is similar to  VW~LMi, but its  outer period is
longer, 2.9 yr; the inner periods are 2.08 and 1.42 days, and all four
masses are comparable,  from 2 to 3 \msun.   The compact 2+2 quadruple
with  the second-longest  outer period  of 432  days  is TIC~454140642
(04191+0054),     a     doubly-eclipsing     star    discovered     by
\citet{Kostov2021}. Its inner periods are  13.6 and 10.4 days, and the
masses  of all four stars are  remarkably  similar, from  1.1 to  1.2
\msun. The system is not young (estimated age 1.9 Gyr).

Modern large-scale  photometric surveys such  as OGLE revealed  a vast
number of  eclipsing binaries. Some of  those turned out  to be doubly
eclipsing, being  in fact  2+2 quadruples or  higher-order hierarchies
\citep{Zasche2019}.   Double   eclipses  are  more   likely  when  the
subsystems  are  mutually  aligned,   which  is  expected  in  compact
hierarchies  like   VW  LMi.  Indeed,  in   some    doubly-eclipsing
quadruples it was possible to detect  motion in the outer orbit by the
anti-correlated   variations  of  the   eclipse  time.   For  example,
OGLE~BLG-ECL-145467 (17521$-$2920) has the  outer period of 4.2 yr and
the inner  periods of  4.9 and  3.9 days. One  of the  most intriguing
findings of  this work  is the hint  on mutual resonances  between the
periods of the inner subsystems: the 3:2  ratio of inner periods is
 frequent  and the  1:2  ratio  seems to  be  rare.  This can  be
explained by the interaction of both subsystems with their outer orbit,
which  is  possible  only  in  compact  configurations.   Yet  another
necessary  condition for such  resonances is  migration of  the orbits
\citep{Tremaine2020}.

Proposing a  formation scenario  of compact 2+2  quadruples of  VW LMi
type  is   a  challenge.   Some  of  their   properties  (compactness,
coplanarity,   similar   masses)   point   to   the   accretion-driven
migration.  However, the  DI+DI  scenario for  compact triples,  where
companions  form  and migrate  sequentially,  does  not  work for  2+2
quadruples. It seems that relatively wide 2+2 quadruples formed by one
of the mechanisms outlined above (CF+CF, CF+DI2, collision) evolved to
their   present   compact   configurations  through   accretion-driven
migration.   This  scenario  predicts   that  {\em  all}  compact  2+2
quadruples must have  large inner mass ratios, as  observed in VW~LMi,
TIC~454140642, V994~Her, and OGLE~BLG-ECL-145467.  Coplanarity is also
predicted because the gas accreted  by the inner subsystems is aligned
with the outer orbit.

\subsection{Castor-type Quadruples}
\label{sec:Castor}

Castor ($\alpha$~Gem, 07346+3153) is a system containing 6 or 7
stars. Its central binary A,B consists of two similar bright early-A type stars
and it has been measured  since 1778; however, the orbital period of 460 yr is
not yet fully covered. Both components of this visual
binary contain spectroscopic subsystems with periods of 9.2 and 2.9
days and minimum masses of 0.2 and 0.36 \msun, respectively. Small
inner mass ratios distinguish Castor from the majority of 2+2
quadruples where all four components typically have similar masses. 

The architecture  of the  sextuple  system 88~Tau  (04357$+$1010) is  very
similar to that of Castor, but  its inner quadruple has a period of 18
yr and its  inner spectroscopic subsystems have periods  of 7.9 and 3.6
days  and relatively  large mass  ratios (spectral  types from  G2V to
A6V). The interferometric  study by  \citet{Lane2007}  revealed that  the
inner  subsystems in  this quadruple  are not  aligned with  the 18-yr
outer  orbit  and the  mutual  inclinations  $\Phi$  are  $143^\circ$  and
$82^\circ$. In this work, the inner subsystems were not resolved
directly, and the orientation of their orbits was determined from the
wobble in the motion of the 18-yr pair. 

In  hierarchies assembled inside-out,  the inner  subsystems typically
have large mass ratios  resulting from the accretion-driven migration.
The  architecture of  Castor is  quite distinct  and suggests  that it
could form in reverse order  by the CF+DI2 scenario, where an existing
binary A,B experienced an episode  of late accretion burst that formed
the inner subsystems by  disk instability. The accretion episode could
be caused by an encounter with another protostar surrounded by its own
disk that also became unstable and formed a close pair that remained
bound to the inner binary A,B, making it a sextuple system. Although the
proposed   scenario  looks   exotic,  it   can  explain   the  unusual
architecture of Castor-type hierarchies.

Recently,  another  sextuple system  with  an architecture  resembling
Castor  has  been  discovered  by  \citet{Powell2021}.   This  is  the
triply-eclipsing star TIC 168789840 (04141$-$3155).  It contains three
eclipsing subsystems  with periods  of 1.3, 1.5,  and 8 days;  all six
sets of  eclipses (3 primary and  3 secondary) are  observed.  The two
closest binaries are paired in an  orbit with a period of a few years,
while  the 8-day pair  belongs to  the outermost  component on  a wide
orbit with  a period of $\sim$2  kyr.  The edge-on  orientation of the
three inner orbits could be  a mere coincidence. More likely, however,
the inner  orbits in this  triply-eclipsing hierarchy  are alined  (the same
probabilistic argument was advanced for the doubly-eclipsing systems).
Another  particularity  of  TIC~168789840  consists in  the  remarkable
similarity  of the  component's  masses and  inner  mass ratios.   The
primary components  of all eclipsing subsystems have  masses about 1.4
\msun, and  their secondaries  are all about  0.6 \msun.  It  looks as
though  all three  inner  subsystems  were made  at  the same  factory
according to  the same `blueprint'.   A potential formation  scenario of
this system is  CF+DI2, where a young triple  composed of similar-mass
stars  encounters a  gas  cloud,  and the  accretion  burst forms  
secondary subsystems with similar  properties around each component of
the original triple.

\subsection{Quadruples of 2+2 Hierarchy}
\label{sec:2+2}

Quadruple systems of 2+2 hierarchy (two close binaries on a wide orbit
around each  other) are rather  common. Their fraction  among solar-type
stars  is  4\%,  compared  to  the  1\%  fraction  of  3+1  quadruples
\citep{FG67}. Compact 2+2 quadruples like VW LMi and the Castor-type
quadruples discussed above belong to this group, which contains
hierarchies with a wide range of periods and mass ratios. 

The classical  visual quadruple system  $\epsilon$~Lyr (18443+3940, HR
7051-7054) consists of  four similar stars of spectral  types from F1V
to A4V, arranged  in two pairs with periods of 1800  and 700 yr (these
orbits  are poorly  constrained) at  a large  projected  separation of
11.5$\times$10$^4$ au.  The  similarity of all four masses  and of the
two  inner  periods  is  the  characteristic of  this  family,  called
$\epsilon$~Lyr type for  brevity \citep{Tok2008}.  Other, more compact
members of  this family are  known, e.g. the visual  quadruple FIN~332
(18455+0530, HR~7048), where all four components are similar (spectral
types from A0V to A1V), the  outer projected separation is 432 au, and
the two inner  orbits have periods of 40 and 48  yr; both inner orbits
also have large eccentricities of 0.82 and 0.84 \citep{Tok2020b}.  Yet
another quite  typical example is HIP 43947  (08571$-$2951), where the
two pairs of similar solar-type stars, at 100 au projected separation,
have separations of 3-4 au each and estimated periods of $\sim$100 yr. 

Maximum periods of inner subsystems in a 2+2  quadruple are limited by
the dynamical stability criterion,  hence some correlation between the
two inner periods is expected.   However, the statistical analysis of such
quadruples in  \citep{Tok2008} shows that the similarity  of the inner
periods is stronger  than required for the stability alone,  hinting that the
inner pairs  have not formed  independently of each  other.  Likewise,
the two inner mass ratios in 2+2 quadruples appear to be correlated.

The $\epsilon$~Lyr  type 2+2  quadruples resemble products  of cascade
hierarchical fragmentation  driven by the conservation  of the angular
momentum   in   the   orbits   of   the   subsystems,   envisioned   by
\citet{Bodenheimer1978}.    The  relatively   wide  outer   and  inner
separations (above  the opacity  limit for fragmentation)  and similar
masses  of the  components  match the  predictions  of this  scenario.
Nevertheless, the approximate  coplanarity of orbits implicit for this mechanism does not
hold.  Although the  sample of resolved visual 2+2 quadruples  with known sense
of revolution  is rather small,  the numbers of  apparently co-rotating
and counter-rotating  inner pairs are similar,  demonstrating the lack
of mutual alignment  between  the   inner  subsystems.  However,  the  outer
separations in  this sample  are larger than  $\sim$1000 au,  and such
wide  triples  also  have  misaligned  orbits.  In  the  more  compact
quadruple FIN~332 the two inner  orbits could be mutually aligned, but
the outer pair rotates in  the opposite sense, excluding  alignment
of the whole system.

The $\epsilon$~Lyr  system is wide, unlike the  compact 2+2 quadruples
presented above, but  both groups have one common  feature, namely the
similarity of masses. However, 2+2 quadruples with unequal masses are also
quite  common.   A relevant  example  is  HIP~12548 (02415$-$7128),  a
classical visual binary with an outer period of 100 yr and a semimajor
axis  of $0.6''$.  Its two  solar-type  components A  and B,  resolved
visually   or  interferometrically,   are  never   separated   by  the
seeing-limited spectroscopy.   Existence of a  spectroscopic subsystem
was   suspected  from   the  occasional   doubling  of   the  spectral
lines. Regular  monitoring revealed that  each component is in  fact a
spectroscopic binary. Their periods  are relatively long, $\sim$2010 and 110
days, and the inner mass  ratios are small, $\sim$0.3 (Tokovinin 2021,
in  preparation).  Such  quadruples  are very  difficult to  discover.
Other  visual  binaries  where  each  component  has  a  spectroscopic
subsystem are known, e.g. HIP 41171 (08240$-$1548) with an outer period
of $\sim$900 yr and the inner periods of 25.4 and $\sim$960 days. The masses
of  all components  of HIP~41171  are comparable  and in  some orbital
phases the spectrum contains  four distinct systems of spectral lines,
although most of the time the lines are blended.

Summarizing, 2+2 quadruples have a wide range of parameters, from wide
configurations of $\epsilon$ Lyr type consisting of similar-mass stars
to  the  most  compact  known  quadruple  VW~LMi,  also  with  similar
masses. Most 2+2 quadruples are  situated in the middle of this range,
and their masses  can be quite diverse. Such  quadruples are difficult
to discover, especially when  their secondary components are faint and
do  not have  signatures in  the spectrum.  One notable  exception are
doubly-eclipsing  systems. The  diversity of  2+2  quadruples suggests
that they could be formed by a variety of mechanisms.

\subsection{Misaligned Hierarchies}
\label{sec:mis}

The  trend to  mutual orbit  alignment weakens  with  increasing outer
separation.  However,  misaligned  hierarchies  exist  even  at  small
separations. Several such systems are well documented. 

A typical example of a misaligned triple is $\zeta$~Aqr (22288$-$0001,
HR 8559+8558)  \citep{Tok2016}. This is  a bright visual  binary known
since  1777. A  wobble  in  its motion  on  the 540-yr  outer orbit  was
detected in 1955, and the  inner subsystem Aa,Ab was directly resolved
in  2009; its 26-yr  orbit is  now well  defined. The  most remarkable
feature of this  triple is the large inner  eccentricity, $e_{\rm in}=
0.87$, and the large mutual  inclination of $140^\circ$: the inner and
outer  pairs  are definitely  counter-rotating.  The  period ratio  of
20.8$\pm$0.6 is  small  and, considering the outer eccentricity of
0.42, this triple is located near the limit of dynamical stability
(eq.~\ref{eq:Mardling}).  The visual  components A  and B  have similar
masses  of 1.4  \msun  (an outer  twin),  but the  inner secondary  is
substantially less massive, 0.6 \msun.

Although the modest  period ratio and the outer  semimajor axis of 100
au  roughly  match  parameters  of  planar  visual triples,  other
characteristics  of $\zeta$~Aqr  are quite  distinct. The  large inner
eccentricity and  the counter-rotation resemble  products of dynamical
interactions.   Usually, the  least  massive star  is  ejected from  a
dynamically unstable triple, but  we may envision a reverse process,
namely a capture.  A low-mass star can encounter  the existing binary,
interact with  it dynamically, and become captured.   The mechanics of
point  masses  is  time-reversible  and allows  capture  (opposite  of
ejection),  given the  appropriate  initial conditions.  A capture  is
probably more  likely in an encounter of  two binaries because  in this  case one
star  can  be ejected,  removing  the  excess  of energy  and  angular
momentum that is  needed for a capture (in  a triple-star capture, the
energy and momentum  are absorbed by the outer  binary).  Equal masses
of the  main components A and  B speak for their  common origin, while
the  low-mass  inner  companion  could  be  formed  independently  and
captured later.

This example  is by no means  unique.  Another very  similar system is
02460$-$0457 (HD  17251).  This is  a $1''$ long-period  visual binary
composed of two solar-type stars  (masses 1.4 and 1.0 \msun). A wobble
has been suspected in its motion, and the faint inner companion has been
eventually directly  detected in 2016.  Its  orbit, still preliminary,
has a period of 38 yr,  an eccentricity of 0.6, and the rotation sense
opposite to the outer binary; the  mass of the inner companion is only
0.44 \msun.  The resemblance to  $\zeta$~Aqr is quite obvious.  In yet
another nearby triple 20396+0458  (HIP 101955), the mutual inclination
is $65^\circ$, so the orbits  are closer to being perpendicular rather
than   coplanar.   Not   surprisingly,  the   inner   eccentricity  is
substantial,  0.69, and  the LK  cycles are  certainly going  on.  The
periods are 38.7 and 2.51 yr (ratio 15.4), and the masses of all three
stars are comparable, between 0.6 and 0.7 \msun.

The statistical  study of mutual  orbit orientation in  visual triples
\citep{Tok2017} hints at a decreased alignment with increasing stellar
mass. Massive stars  form in dense environments and  are more prone to
dynamical interactions with their cluster neighbors and with their own
companions.   Moreover, assembly  of  massive stars  implies   strong
accretion which favors companion formation by DI and their fast inward
migration.   Migrating  outer   companions   de-stabilize  the   inner
subsystems,  leading to  ejections  (runaway stars),  and leave  behind
hierarchies with  misaligned and  eccentric orbits.  The  most compact
known triple,  $\lambda$~Tau (04007+1229, HR  1239) has a  B3V primary
component; its periods are 33.07  and 3.953 days (ratio 8.37, near the
limit of  dynamical stability). The  inner pair is eclipsing,  and the
inclination of the 33-day  pair is estimated at $65^\circ$, suggesting
misaligned orbits despite the short outer period.  A better documented
example of a misaligned massive triple is $\sigma$~Ori (05387$-$0236,
HR~1931),    an    O9.5V     member    of    the    Orion    Trapezium
cluster. \citet{Schaefer2016} resolved the inner 143-day spectroscopic
subsystem with  an eccentric ($e=0.77$) orbit and  determined that its
inclination  to the  outer visual  orbit of  157 yr  period  is either
$130^\circ$ or $114^\circ$, so the subsystems counter-rotate.

\subsection{High-Multiplicity Systems}
\label{sec:N}

\begin{figure}
\includegraphics[width=8.5 cm]{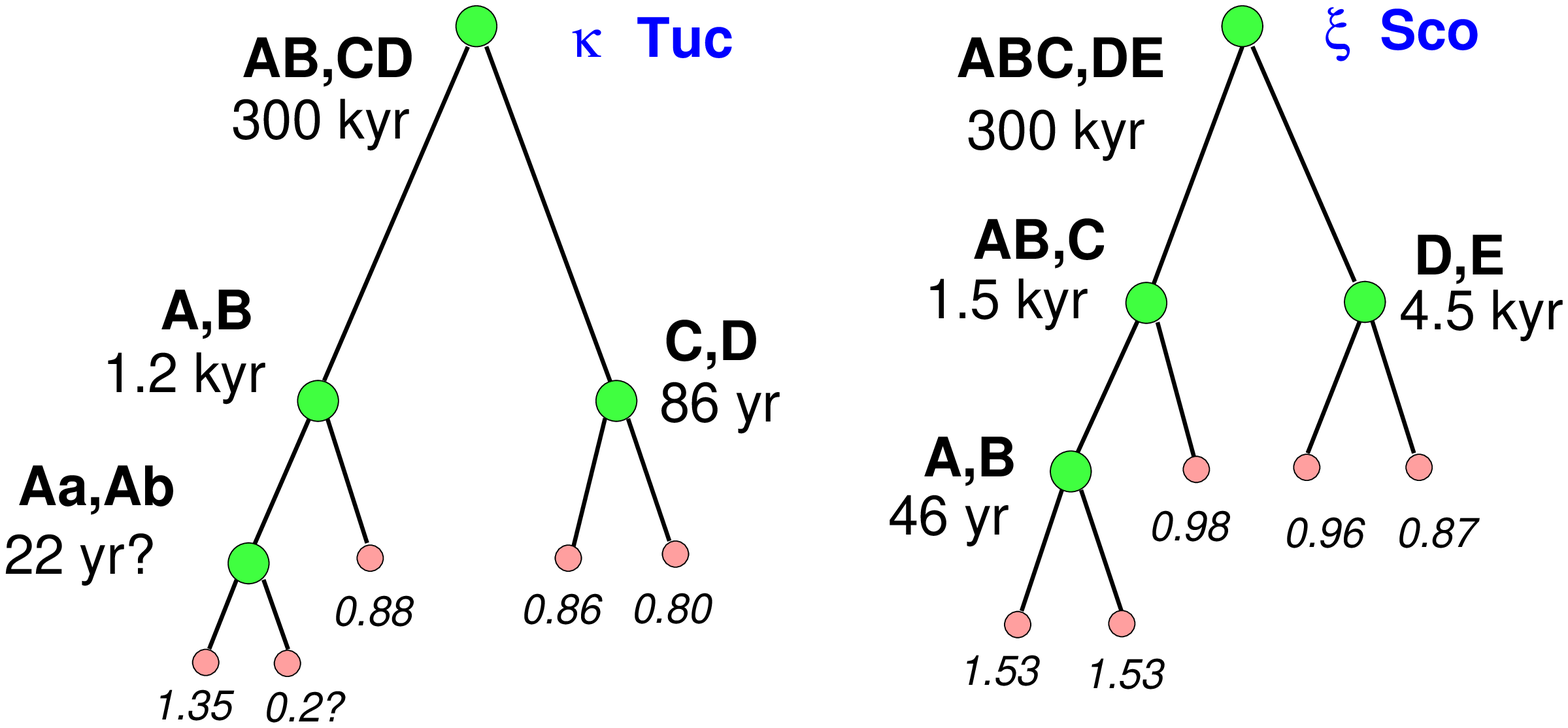}
\caption{Hierarchical structure of two wide nearby quintuple systems
  $\kappa$~Tuc and $\xi$~Sco. Periods of the subsystems (green
  circles) and masses of the stars (pink circles) are indicated. 
\label{fig:KsiSco} }
\end{figure}

Hierarchies containing  5, 6,  or 7 stars  can be  collectively called
{\em  large-N} systems. Architecture  of one  such system,  65~UMa, is
illustrated   in    Figure~\ref{fig:65UMa}.    Castor   ($\alpha$~Gem,
07346+3153) contains  6 or 7  components.  Its central  quadruple AB,
discussed above, is  paired with the twin eclipsing  binary C = YY~Gem
at projected separation  of 1100 au.  The period of  this pair Ca,Cb is 0.81
days.  \citet{Wolf2018} discovered eclipse time variation with a 54 yr
period that  could be caused by  a substellar companion  of 0.05 \msun
associated with C.  If this  companion is real, Castor is a septuple
system.  It has only three  hierarchical levels, allowing a maximum of
$2^3 = 8$  components. In this system, almost  all available slots are
filled, which is  atypical.

A large-N hierarchy can  be decomposed into simpler constituents. For
example, the  Castor system is a  combination of the 2+2  quadruple and a
triple. The large number of components is normally associated with the
large  range  of  separations:   the  outer  systems  in   large-N
hierarchies are wide, and the inner subsystems are close, as in Castor.

Two nearby (within  30 pc) quintuples that defy  the general trend and
contain   only  relatively   wide   subsystems,  namely   $\kappa$~Tuc
(01158$-$6853)   and  $\xi$~Sco   (16044$-$1122),  are   discussed  by
\citet{Tok2020a}  and illustrated  in  Figure~\ref{fig:KsiSco}.  Their
outer periods  are $\sim$300 kyr (projected  separations $\sim$8 kau),
and all  inner periods  are longer than  20 yr.  Both  hierarchies are
$\sim$2  Gyr old  and, naturally,  are not  associated with  any young
kinematic  group.  The wide  outer subsystems  indicate an  absence of
strong dynamical interactions  with neighbors, suggesting formation in
a  low-density environment  like Taurus.   The age  of  these systems
speaks  for  their stability,  ruling  out  strong internal  dynamical
interactions.  In harmony with  the wide outer separations, the orbits
of the  inner visual subsystems  appear to be oriented  randomly (some
pairs revolve in  opposite directions).  Masses of the  stars in these
hierarchies are  similar (from 0.8  to 1.5 \msun)  and do not  match a
random selection from the IMF.  The formation scenario of such systems
proposed in  \citep{Tok2020a} is a filament  fragmentation in relative
isolation (see above).  The similarity of masses is possibly explained
by the prolonged accretion of gas on to the systems; however, for some
unknown reason the accretion has not shrunk the inner orbits.

Given that the multiplicity is  a strong function of stellar mass, one
might expect  that large-N hierarchies are rare  among low-mass stars;
at least, such is the prediction of the IMM statistical model (section
\ref{sec:stat}). Yet,  low-mass quintuples are found  in the immediate
vicinity  of  the  Sun,  casting  doubt on  their  rareness.   GJ~2069
(08316+1924, HIP  41824, parallax 60 mas) contains  five similar stars
of spectral  types from M4V to  M3.5V arranged in  a 3+2 architecture:
two visual binaries at 10.4$''$ ~from each other, with one of those
containing  a 2.8-day  spectroscopic and  eclipsing  subsystem CU~Cnc.
Even  more  interesting is  the  quintuple  GJ~644 (16555$-$0820,  HIP
81817, Wolf 630)  located at 6 pc from the  Sun. Its innermost compact
triple  composed  of  M3V  dwarfs  (periods  1.7 yr  and  3  days)  is
accompanied  by an M4V  dwarf at  21$''$ and  another M7V  dwarf at
232$''$, making it a rare 4-tier hierarchy.

\subsection{Young Hierarchies}
\label{sec:young}

In the standard  paradigm of low-mass star formation,  the envelope is
accreted  in  less  than  $10^5$  yr,  so  the  chances  of  observing
multiple-star formation directly are small.  On the other hand, in
the hierarchical  collapse framework the accretion  and mass assembly
continue  for  1--10  Myr  and  is  still  ongoing   in  the  nearby
star-forming regions (SFRs) like Taurus-Auriga and Orion. Formation and early
evolution  of hierarchical systems  is therefore  directly observable.
Study of young hierarchies helps to develop and test the formation scenarios.

The prototype  of pre-main  sequence (PMS) stars,  T~Tau (04220+1932),
can also be considered as  a prototype of typical triple systems.  Its
heavily obscured southern component  S, at 100 au projected separation
from the northern  star N, is itself  a binary with a period  of 27 yr
and masses  of 2.1 and 0.5  \msun. The binary Sa,Sb  is accreting from
the  circumbinary  disk  and  drives   an  outflow;  star  N  is  also
accreting. The kinematics of the  gas and stars is complex, indicating
that this system  is misaligned and `in disarray',  using the words of
\citet{Kasper2020}. Most of the  mass is concentrated in the component
S hosting the inner  subsystem, suggestive of the inside-out formation
scenario by the DI+CF or CF+CF  mechanisms.  The fact that star N also
accretes speaks against its dynamical ejection; more likely, it formed
in the same cloud, approached S, and was captured. Continued accretion
and dynamical evolution  will shrink both inner and  outer orbits, and
in the  end T~Tau might become  a typical triple system  with an outer
period on the order of  a few centuries, its primary component hosting
a close subsystem Sa,Sb.

Interesting insights  on the  multiple-star formation are  provided by
the PMS  triple GW Ori  (05291+1152) located in the  $\lambda$~Ori SFR
with an age  of $\sim$1 Myr.  The inner binary  consists of stars with
masses of 2.5  and 1.4 \msun on a nearly circular  orbit with a period
of 241.6 days. The 1.4 \msun tertiary component C has a period of 11.5
yr,  and  the  eccentricity of  its  orbit  is  0.38. The  orbits  are
well-constrained,    their    mutual    inclination   is    $14^\circ$
\citep{Kraus2020}.   This  system   is  actively  accreting  from  the
surrounding gas  which has a  remarkable structure. The  dust emission
comes mostly from  two coplanar outer rings with the  radii of 180 and
350   au  and  from   the  tilted   inner  ring   of  43   au  radius.
Interferometric observations in the mm range and high-contrast imaging
in the  IR revealed the 3D  structure of these  rings in unprecedented
detail \citep{Kraus2020}. All rings rotate  in the same sense, but the
inner ring is inclined to the outer orbit of the triple by $39^\circ$.
Hydrodynamic simulations can reproduce  the system's geometry and show
that the inner ring was torn  from the disk by its dynamic interaction
with the  triple; for the same  reason the outer disk  is warped. Like
T~Tau, this object proves that  mass assembly can last for $\sim$1 Myr
and  the angular  momentum  of  the infalling  gas  can be  misaligned
relative to the stellar system.
 
When  several   stars  form   in  close  proximity,   their  dynamical
interactions are  inevitable. Although the overall  stellar density in
the Taurus-Auriga SFR is low, the association is highly structured and
contains    compact   dense    groups   of    stars.    According   to
\citet{Menard2020},  the triple  system  UX~Tau (04301+1814)  presents
evidence of a recent fly-by of another star. Its main components A and
B  are separated by  $5.7''$ (825  au); B  is a  $0.1''$ binary,  A is
surrounded  by a  disk. There  is another  star C  at $2.7''$  from A.
Comparable  separations  of AB  and  AC  imply dynamical  instability.
However, M\'enard  et al.  show that  star C is  likely a  fly-by.  It
interacted recently with the disk of A, leaving a typical signature in
the form of spirals, and captured  some gas into its own small disk; a
tenuous gas  bridge still connects A  and C. The  authors cite another
examples  of  fly-bys of  PMS  stars  evidenced  by the  residual  gas
structures such as bridges, tidal tails, and spirals.

A    nascent    trapezium     system    has    been    discussed    by
\citet{Reipurth2021}. This is IRAS 05417+0907 in the $\lambda$ Ori SFR
associated with the Herbig-Haro flow HH 175.  This object contains six
sources  at comparable  separations on  the  order of  a few  thousand
au. One of those stars is  immersed in a dense envelope and drives the
outflow. The authors suggest that this is a binary formed by dynamical
interaction in  the trapezium and that the  associated accretion burst
created  the outflow;  its dynamical  age is  6\,000 yr.  An eccentric
binary continues  to accrete gas, preferentially  near periastron, and
evolves to closer separation. Some Herbig-Haro jets (e.g. HH 111) have
quasi-periodic knots that could be produced by accretion bursts driven
by   the   inner   evolving   binary   of   $\sim$10   au   separation
\citep{Reipurth2000}. Apart from the N-body dynamics, binaries driving
the  jets can  result from  gas-assisted captures,  as happens  in the
cluster simulations.

GG~Tau (04325+1732) is a  famous PMS hierarchy. Its components A
and  B,  separated  by  10$''$  (1400 au),  are  close  binaries  with
circumbinary disks,  and AB is surrounded by  a circum-quadruple disk,
earning this object the nickname `ring world'. Moreover, the component
Ab is also a tight $0.03''$  pair, making this system a quintuple. The
masses of these stars range from  0.2 to 0.65 \msun. A recent paper by
\citet{Keppler2020} explores  the structure  of the disk  around Aa,Ab
(cavity,  filaments,  shadows,  accretion  streams), trying  to  infer
orientations of the unresolved circumstellar disks.

The  2+2  PMS  quadruples  are  represented by  the  system  HD  98800
(11221$-$2447),  also known  as TWA~4  because it  belongs to  the TWA
association (age  $\sim$10 Myr). The  outer system A,B is  a classical
visual binary with a period  of 200-300\,yr and a moderate eccentricity of
0.4;  the  orbit of  A,B  is highly  inclined.   Both  stars are  also
spectroscopic   binaries   with  periods   of   262   and  315   days,
respectively. Moreover, the  double-lined subsystem Ba,Bb was resolved
using Keck interferometer, allowing \citet{Boden2005} to determine the
mutual inclination and the masses (0.7  and 0.6 \msun). B is surrounded by
a  dusty and   gaseous circumbinary  disk responsible  for the
large IR  excess, while A  has no disk.   The disk around  B, directly
imaged by ALMA  \citep{Kennedy2019}, is a ring of 3.5  au radius and 2
au  width,  oriented   almost  `face-on'  (inclination 
154$^\circ$). Hence, this disk is perpendicular to the orbits of Ba,Bb
and  A,B.   The  subsystem  Aa,Ab  also  has  been  recently  resolved
interferometrically   \citep{ZF2021}.

A nascent  system of three protostars  L1448 IRS3B in  the Perseus SFR
(age  less   than  150  kyr)   has  been  discovered  and   studied  by
\citet{Tobin2016}. The inner binary consists of two solar-mass stars A
and B  at projected separation of 61  au. It is immersed  in a massive
(0.3 \msun) gaseous  disk with a spiral structure,  and another star C
is forming in this disk at  projected separation of 183 au. Star C has
a low mass, but looks brighter than A and B owing to the faster accretion.
The authors present this system as a classical case of the DI+DI
formation scenario; the massive disk surrounding A and B is
gravitationally unstable in the region where C is forming.

\section{Summary and Discussion}
\label{sec:sum}

Many observational and theoretical  works mentioned in this review are
quite recent,  evidencing that architecture of  stellar hierarchies is
presently a very active research topic.  The interest is stimulated by
the synergy  with exoplanets,  where  hierarchies help  to understand  the
origin  and evolution  of  the  angular momentum  of  stars and  their
systems and inform us on the dynamics of gas from which planets
will  form. New  observational techniques  such as  interferometry and
accurate   photometry and astrometry from   space  also   stimulate
this  research. Our concepts related to formation of stars and their
systems evolve in response. These shifting paradigms are summarized
below, followed by an outlook of observational and theoretical
progress expected in the near-term future.

The main conclusion from this  review is the diversity of hierarchical
systems and  the inferred diversity of their  formation scenarios.  The
three  main  ingredients  of  this  process  are  {\em  fragmentation,
  accretion, and  dynamics.}  The scenarios proposed here  are based on
general  considerations  (e.g.    hierarchical  collapse  and  angular
momentum   extraction)  and  recent   simulations,  but   they  remain
tentative.

\subsection{Evolution of Concepts}

{\em 1.  Star formation is  a process, not  an event.} Early  works on
star formation  considered collapse of isolated clouds  happening on a
relatively   short   free-fall   time   scale  of   $\sim   10^4$   yr
\citep{Larson1972}. Now  we know that  assembly of stellar  masses, as
well as formation and evolution of multiple systems, extends over 1-10
Myr; this time scale is defined  by the size of the largest collapsing
structures,  molecular clouds  and clusters  \citep{Vasquez2019}. Star
formation  proceeding during  this time  is  now  likened  to a  `conveyor  belt'
\citep{Krumholz2020},  and  this notion  helps  to  model the  IMF/SMF
\citep{Clark2021}. The  architecture of stellar systems  is defined by
their early evolution as much as by the elementary mechanisms of their
formation; in fact, systems formed  by different mechanisms such as DI
and  CF may  end up  in similar  configurations. Complex  evolution of
stellar  systems   in  a  young  collapsing   cluster  is  beautifully
illustrated by the simulations of \citet{Bate2019}.

{\em 2.   The critical role of  gas} in the formation  and dynamics of
stellar  systems  is  incontestable.  Accretion-driven  migration
forms    close    binaries    during    the    mass-assembly    period
\citep{TokMoe2020}.  Dissipation of kinetic  energy in the gas is also
relevant for  the formation of wide systems  by disk-assisted capture.
In the past, multiple systems were often treated simply as gravitating
point masses.  Only at  close separations, the point-mass dynamics was
modified  by considering  the  finite stellar  size  and tidal  forces
\citep{Naoz2016}.  At large separations, interactions with neighboring
stars  destroy  wide pairs  by  the  so-called `dynamical  processing'
\citep{DK13}.  However,  the N-body dynamics alone fails  to match the
architecture of  real hierarchical systems, as  demonstrated, e.g., by
the  simulations of  \citet{Sterzik2002}.  
Accreting   stars  and   their  systems   evolve jointly and  self-consistently
\citep{Bate2019}. Early works where a `dynamical processing' of binary
population in a  cluster was studied by assuming  some {\em ab initio}
binaries with universal properties  slowly lose their relevance.  This
said, dynamical evolution of  mature hierarchies (after gas dispersal)
leads     to     several     astrophysically    important     outcomes
\citep{Hamers2020,Toonen2020}, while  dynamical interactions during or
shortly after formation of  hierarchies certainly take place and leave
their  imprints  such  as  misaligned hierarchies  or  very  eccentric
orbits.

{\em  3.    Non-universality}  of  star  formation   in  general,  and
multiplicity  in particular, is  not yet  fully appreciated.   The old
concept of a  universal IMF was traditionally extended  in the past to
assume a  universal multiplicity.   High multiplicity  in the  SFRs like
Taurus,  compared  to the  field,  has  been  attributed to  the dynamical
processing in clusters.   Now we know that the  metallicity of the gas
affects  its  fragmentation  scale  and  leads  to  a  strong  inverse
dependence    of   the    close-binary    fraction   on    metallicity
\citep{Moe2019}. Conversely, the  fraction of wide binaries positively
correlates   with  metallicity  \citep{Hwang2021}.    At  intermediate
separations  of  300-3000  au,  the  fraction  of  binaries  in  loose
associations   is   twice   lager    than   in   the   open   clusters
\citep{Deacon2020}.   This  difference  highlights the  dependence  of
multiplicity on the density of  the formation environment (it would be
difficult to explain the difference by dynamical processing, which is effective
only  at   larger  separations).   The  notions   of  binary-rich  and
binary-poor  formation  environments  help  us to  understand  certain
aspects of hierarchies mentioned above.

\subsection{Observational Perspective}

Comparing  predictions   of  formation  theories   with  the  observed
statistics of stellar systems has  always been the main way of testing
the  theory and  gaining  new insights.   However,  the population  of
stellar  systems  in the  field  results  from  a mixture  of  various
formation channels.  Moreover,  statistical properties of binary and
multiple  systems  depend   substantially  on  their
formation  environments, as  emerges  from the  recent studies.   This
consideration somewhat undermines the usefulness of statistics derived
from the  nearby field  population.  

On the  other hand, a reasonable  completeness over the  full range of
periods can be  reached only for the nearby  field.  Specifically, the
multiplicity statistics of the  nearby low-mass stars and brown dwarfs
can be revealing because the  DI formation channel is unlikely and the
accretion-induced evolution of the orbits may be less important. Thus,
the architecture  of low-mass hierarchies is more  directly related to
the CF-based  formation scenarios.  Two nearby  quintuples with stellar
masses of  $\sim$0.4 \msun  are mentioned in  section~\ref{sec:N}. The
triple  LHS~1070 and  the young quadruple with substellar masses
studied by  \citet{Bowler2015} are extreme examples  of such systems.
Discovery and study of more low-mass hierarchies are needed.

The origin of  stellar hierarchies is also elucidated  by the detailed
study of selected `benchmark' systems, linking their properties to the
formation  theories.  This  approach,  most popular  in  the study  of
nascent stars  (section~\ref{sec:young}), can be  usefully extended to
mature   stellar  hierarchies  because   attempts  to   explain  their
architecture  might be challenging  and informative  at the  same time
\citep[e.g.][]{Powell2021}.   Here  I explored  a  hybrid strategy  by
dividing  known hierarchies  into  groups according  to some  relevant
parameters  and  associating  these  groups with  different  formation
scenarios.  This  classification attempt can be  rightly criticized for
being  subjective and  ambiguous.  For  example, Castor  is  a large-N
system,  it contains  a 2+2  inner  quadruple, and  this quadruple  is
attributed  to the special  `Castor family'  presumably formed  by the
CF+DI2  scenario; $\lambda$~Tau  is compact,  but, at  the  same time,
likely misaligned.

Accumulation  of   new  data  on  stellar   hierarchies  is  currently
progressing  at an  increased pace.   The  classical, object-by-object
studies  of  hierarchies  are  nowadays complemented  or  replaced  by
large-scale  surveys from  space,  e.g.  Gaia  \citep{Gaia2} and  TESS
\citep{TESS}, that  detect new hierarchical systems in  large numbers.  The
ground-based  photometric  and   spectroscopic  surveys  also  make  a
substantial   contribution  \citep[e.g.][]{Zasche2019}.   Well-defined
coverage and uniform  detection limits of surveys favor  their use for
statistical studies.   However, the sky surveys  are usually not designed  for the
study of hierarchies and bring  only partial information that helps to
discover  such  systems but  is  often insufficient  for  their study.   For
example, the cadence  of the Gaia RV data  will allow determination of
spectroscopic orbits  in a  restricted range of  periods and  only for
single-lined systems, because splitting blended lines requires a larger
spectral  resolution and  dedicated methods.   Similarly,  the Gaia mission
duration will  restrict the periods of future  Gaia astrometric orbits
to less than a few years.

Combination  of  surveys  with  dedicated  follow-up  observations  of
selected objects is nowadays  the mainstream of observational progress
on multiples.   By screening large  samples, surveys help  to identify
rare systems  that either extend  the explored parameters  space (e.g.
compact and  ultra-planar systems found  in Kepler and TESS)  or offer
prospects  of particularly detailed  study. Follow-up  observations of
these   selected   systems    reveal   their   structure   in   detail
\citep[e.g.][]{Powell2021}.   High-resolution studies  of cold  gas by
ALMA add  a new  dimension to the  classical field of  multiple stars,
especially   promising   for   the   PMS  hierarchies   (see   section
\ref{sec:young}).

The  uniform coverage  and  the  large volume  of  modern surveys  can
partially  compensate for the  insufficient information  on individual
systems  by using statistical  approach, where  the statistics  of the
underlying population are inferred  by modeling the observed parameters
(e.g. astrometric accelerations in  the future Gaia data releases) and
solving  the  inverse problem.  For  example,  thousands of  eclipsing
binaries discovered  in photometric  surveys, coupled with  modeling of
the   observational  selection,  provide   solid  information   on  the
underlying population  of close binaries.  Statistical exploitation of
large surveys in terms of  stellar multiplicity is expected to augment
in the near future.

Many thousands of  new binaries discovered by Gaia  and their accurate
astrometry can contribute substantially to the study of relative orbit
orientation   by   the   indirect   method   of   counting   co-   and
counter-rotating  systems; one  could  probe mutual  inclination as  a
function of mass and age.  Relative motions in wide pairs also contain
statistical infromation  on their eccentricities, allowing  to test the
unfolding    mechanism    of    \citet{Reipurth2012}.    Intriguingly,
\citet{Shatsky2001} found  that wide pairs  containing subsystems tend
to  have  less  eccentric  orbits  than  simple  binaries,  in  direct
contradiction with the predictions of unfolding.

\subsection{Theoretical Perspective}

The  way   forward  from  the  current   tentative  classification  of
hierarchies  based on  the  biased MSC  sample  to a  more robust  and
informative connection  between architecture of  hierarchies and their
formation  mechanisms   is,  as  usual,  offered   by  combination  of
theoretical  advances with  new observations.  I am  not an  expert in the
star formation theory and can only hope  for more quantitative and predictive studies
of  different formation  channels.  The  inherently chaotic  nature of
star formation  and the diversity  of initial conditions  and relevant
processes  are   natural  obstacles  in  this   regard.  Although  all
simulations  are  informative,  the  greatest return  is  expected  by
simulating many  random cases  followed by statistical  analysis of
the results. In this way, predictions of the formation scenarios can be
sharpened   and   quantified,    allowing   better   comparison   with
observations. Establishing the  relative role  of different formation  channels and
quantitative  predictions  of  their  outcomes  is one  of  the  major
outstanding  issues. 

Formation of close binaries and their relation to higher-order
multiples is still an open question. Although the crude modeling by
\citet{TokMoe2020} and the analysis by \citet{Moe2018}  highlight
accretion as the main agent of orbital migration, many details are
still obscure: Is migration in triples always associated with orbit alignment
or not? Can simulations reproduce the formation of compact 2+2
hierarchies?

Even  if in  the future  the goal  of  building a
predictive theory of multiple-star formation will prove elusive (as it
is now), every step in  this direction will bring new valuable insights on
the formation of stars and planets.


\begin{acknowledgments} 

I appreciate comments on the draft of this review provided by N.~Shatsky and S.~Rappaport.
This work  used the  SIMBAD service operated  by Centre  des Donn\'ees
Stellaires  (Strasbourg, France),  bibliographic  references from  the
Astrophysics Data  System maintained  by SAO/NASA, and  the Washington
Double Star  Catalog maintained  at USNO. 

\end{acknowledgments} 


\bibliographystyle{apj}
\bibliography{mult21.bib}

\begin{thebibliography}{}
\expandafter\ifx\csname natexlab\endcsname\relax\def\natexlab#1{#1}\fi

\bibitem[{{Alonso} {et~al.}(2015){Alonso}, {Deeg}, {Hoyer}, {Lodieu}, {Palle},
  \& {Sanchis-Ojeda}}]{Alonso2015}
{Alonso}, R., {Deeg}, H.~J., {Hoyer}, S., {et~al.} 2015, \aap, 584, L8

\bibitem[{{Anosova}(1986)}]{Anosova1986}
{Anosova}, J.~P. 1986, \apss, 124, 217

\bibitem[{{Antognini} \& {Thompson}(2016)}]{Antognini2016}
{Antognini}, J. M.~O., \& {Thompson}, T.~A. 2016, \mnras, 456, 4219

\bibitem[{{Bate}(2019)}]{Bate2019}
{Bate}, M.~R. 2019, \mnras, 484, 2341

\bibitem[{{Batten}(1973)}]{Batten1973}
{Batten}, A.~H. 1973, {Binary and multiple systems of stars} (Pergamon Press:
  Oxford)

\bibitem[{{Boden} {et~al.}(2005){Boden}, {Sargent}, {Akeson}, {Carpenter},
  {Torres}, {Latham}, {Soderblom}, {Nelan}, {Franz}, \&
  {Wasserman}}]{Boden2005}
{Boden}, A.~F., {Sargent}, A.~I., {Akeson}, R.~L., {et~al.} 2005, \apj, 635,
  442

\bibitem[{{Bodenheimer}(1978)}]{Bodenheimer1978}
{Bodenheimer}, P. 1978, \apj, 224, 488

\bibitem[{{Bonnell} \& {Bastien}(1993)}]{Bonnell1993}
{Bonnell}, I., \& {Bastien}, P. 1993, \apj, 406, 614

\bibitem[{{Borkovits} {et~al.}(2016){Borkovits}, {Hajdu}, {Sztakovics},
  {Rappaport}, {Levine}, {B{\'\i}r{\'o}}, \& {Klagyivik}}]{Borkovits2016}
{Borkovits}, T., {Hajdu}, T., {Sztakovics}, J., {et~al.} 2016, \mnras, 455,
  4136

\bibitem[{{Borkovits} {et~al.}(2019){Borkovits}, {Sperauskas}, {Tokovinin},
  {Latham}, {Cs{\'a}nyi}, {Hajdu}, \& {Moln{\'a}r}}]{Borkovits2019}
{Borkovits}, T., {Sperauskas}, J., {Tokovinin}, A., {et~al.} 2019, \mnras, 487,
  4631

\bibitem[{{Borkovits} {et~al.}(2020){Borkovits}, {Rappaport}, {Tan},
  {Gagliano}, {Jacobs}, {Huang}, {Mitnyan}, {Hambsch}, {Kaye}, {Maxted},
  {P{\'a}l}, \& {Schmitt}}]{Borkovits2020}
{Borkovits}, T., {Rappaport}, S.~A., {Tan}, T.~G., {et~al.} 2020, \mnras, 496,
  4624

\bibitem[{{Bowler} \& {Hillenbrand}(2015)}]{Bowler2015}
{Bowler}, B.~P., \& {Hillenbrand}, L.~A. 2015, \apjl, 811, L30

\bibitem[{{Brandt}(2018)}]{Brandt2018}
{Brandt}, T.~D. 2018, \apjs, 239, 31

\bibitem[{{Chambliss}(1992)}]{Chambliss1992}
{Chambliss}, C.~R. 1992, \pasp, 104, 663

\bibitem[{{Clark} \& {Whitworth}(2021)}]{Clark2021}
{Clark}, P.~C., \& {Whitworth}, A.~P. 2021, \mnras, 500, 1697

\bibitem[{{Deacon} \& {Kraus}(2020)}]{Deacon2020}
{Deacon}, N.~R., \& {Kraus}, A.~L. 2020, \mnras, 496, 5176

\bibitem[{{Delgado-Donate} {et~al.}(2003){Delgado-Donate}, {Clarke}, \&
  {Bate}}]{Delgado-Donate2003}
{Delgado-Donate}, E.~J., {Clarke}, C.~J., \& {Bate}, M.~R. 2003, \mnras, 342,
  926

\bibitem[{{Docobo} {et~al.}(2021){Docobo}, {Piccotti}, {Abad}, \&
  {Campo}}]{Docobo2021}
{Docobo}, J.~A., {Piccotti}, L., {Abad}, A., \& {Campo}, P.~P. 2021, \aj, 161,
  43

\bibitem[{{Duch{\^e}ne} \& {Kraus}(2013)}]{DK13}
{Duch{\^e}ne}, G., \& {Kraus}, A. 2013, \araa, 51, 269

\bibitem[{{Dupuy} \& {Liu}(2011)}]{Dupuy2011}
{Dupuy}, T.~J., \& {Liu}, M.~C. 2011, \apj, 733, 122

\bibitem[{{Duquennoy} \& {Mayor}(1991)}]{DM91}
{Duquennoy}, A., \& {Mayor}, M. 1991, \aap, 500, 337

\bibitem[{{Eggleton}(2006)}]{Eggleton2006}
{Eggleton}, P. 2006, {Evolutionary Processes in Binary and Multiple Stars}
  (Cambridge University Press)

\bibitem[{{Eggleton} \& {Kisseleva-Eggleton}(2006)}]{KCTF}
{Eggleton}, P.~P., \& {Kisseleva-Eggleton}, L. 2006, \apss, 304, 75

\bibitem[{{Eggleton} \& {Tokovinin}(2008)}]{Eggleton2008}
{Eggleton}, P.~P., \& {Tokovinin}, A.~A. 2008, \mnras, 389, 869

\bibitem[{{Fabrycky} \& {Tremaine}(2007)}]{Fabrycky2007}
{Fabrycky}, D., \& {Tremaine}, S. 2007, \apj, 669, 1298

\bibitem[{{Fekel}(1981)}]{Fekel1981}
{Fekel}, F.~C., J. 1981, \apj, 246, 879

\bibitem[{{Gaia Collaboration} {et~al.}(2018){Gaia Collaboration}, {Brown},
  {Vallenari}, {Prusti}, {de Bruijne}, {Mignard}, \& {et al.}}]{Gaia2}
{Gaia Collaboration}, {Brown}, A.~G.~A., {Vallenari}, A., {et~al.} 2018, \aap,
  616, A1

\bibitem[{{Hamers}(2020)}]{Hamers2020}
{Hamers}, A.~S. 2020, \mnras, 494, 5298

\bibitem[{{He} {et~al.}(2020){He}, {Ford}, {Ragozzine}, \& {Carrera}}]{He2020}
{He}, M.~Y., {Ford}, E.~B., {Ragozzine}, D., \& {Carrera}, D. 2020, \aj, 160,
  276

\bibitem[{{Heath} \& {Nixon}(2020)}]{Heath2020}
{Heath}, R.~M., \& {Nixon}, C.~J. 2020, \aap, 641, A64

\bibitem[{{Hirsch} {et~al.}(2021){Hirsch}, {Rosenthal}, {Fulton}, {Howard},
  {Ciardi}, {Marcy}, {Nielsen}, {Petigura}, {de Rosa}, {Isaacson}, {Weiss},
  {Sinukoff}, \& {Macintosh}}]{Hirsch2021}
{Hirsch}, L.~A., {Rosenthal}, L., {Fulton}, B.~J., {et~al.} 2021, \aj, 161, 134

\bibitem[{{Hwang} {et~al.}(2020){Hwang}, {Hamer}, {Zakamska}, \&
  {Schlaufman}}]{Hwang2020}
{Hwang}, H.-C., {Hamer}, J.~H., {Zakamska}, N.~L., \& {Schlaufman}, K.~C. 2020,
  \mnras, 497, 2250

\bibitem[{{Hwang} {et~al.}(2021){Hwang}, {Ting}, {Schlaufman}, {Zakamska}, \&
  {Wyse}}]{Hwang2021}
{Hwang}, H.-C., {Ting}, Y.-S., {Schlaufman}, K.~C., {Zakamska}, N.~L., \&
  {Wyse}, R. F.~G. 2021, \mnras, 501, 4329

\bibitem[{{Joncour} {et~al.}(2017){Joncour}, {Duch{\^e}ne}, \&
  {Moraux}}]{Joncour2017}
{Joncour}, I., {Duch{\^e}ne}, G., \& {Moraux}, E. 2017, \aap, 599, A14

\bibitem[{{Kamann} {et~al.}(2020){Kamann}, {Giesers}, {Bastian}, {Brinchmann},
  {Dreizler}, {G{\"o}ttgens}, {Husser}, {Latour}, {Weilbacher}, \&
  {Wisotzki}}]{Kamann2020}
{Kamann}, S., {Giesers}, B., {Bastian}, N., {et~al.} 2020, \aap, 635, A65

\bibitem[{{Kasper} {et~al.}(2020){Kasper}, {Santhakumari}, {Herbst}, {van
  Boekel}, {Menard}, {Gratton}, {van Holstein}, {Langlois}, {Ginski},
  {Boccaletti}, {Benisty}, {de Boer}, {Delorme}, {Desidera}, {Dominik},
  {Hagelberg}, {Henning}, {Heidt}, {K{\"o}hler}, {Mesa}, {Messina}, {Pavlov},
  {Petit}, {Rickman}, {Roux}, {Rigal}, {Vigan}, {Wahhaj}, \&
  {Zurlo}}]{Kasper2020}
{Kasper}, M., {Santhakumari}, K.~K.~R., {Herbst}, T.~M., {et~al.} 2020, \aap,
  644, A114

\bibitem[{{Kennedy} {et~al.}(2019){Kennedy}, {Matr{\`a}}, {Facchini}, {Milli},
  {Pani{\'c}}, {Price}, {Wilner}, {Wyatt}, \& {Yelverton}}]{Kennedy2019}
{Kennedy}, G.~M., {Matr{\`a}}, L., {Facchini}, S., {et~al.} 2019, Nature
  Astronomy, 3, 230

\bibitem[{{Keppler} {et~al.}(2020){Keppler}, {Penzlin}, {Benisty}, {van
  Boekel}, {Henning}, {van Holstein}, {Kley}, {Garufi}, {Ginski}, {Brandner},
  {Bertrang}, {Boccaletti}, {de Boer}, {Bonavita}, {Brown Sevilla}, {Chauvin},
  {Dominik}, {Janson}, {Langlois}, {Lodato}, {Maire}, {M{\'e}nard}, {Pantin},
  {Pinte}, {Stolker}, {Szul{\'a}gyi}, {Thebault}, {Villenave}, {Zurlo},
  {Rabou}, {Feautrier}, {Feldt}, {Madec}, \& {Wildi}}]{Keppler2020}
{Keppler}, M., {Penzlin}, A., {Benisty}, M., {et~al.} 2020, \aap, 639, A62

\bibitem[{{Kervella} {et~al.}(2019){Kervella}, {Arenou}, {Mignard}, \&
  {Th{\'e}venin}}]{Kervella2019}
{Kervella}, P., {Arenou}, F., {Mignard}, F., \& {Th{\'e}venin}, F. 2019, \aap,
  623, A72

\bibitem[{{Kostov} {et~al.}(2021){Kostov}, {Powell}, {Torres}, {Borkovits},
  {Rappaport}, {Tokovinin}, {Zasche}, {Anderson}, {Barclay}, {Berlind},
  {Brown}, {Calkins}, {Collins}, {Collins}, {Conti}, {Esquerdo}, {Hellier},
  {Jensen}, {Kamler}, {Kruse}, {Latham}, {Masek}, {Murgas}, {Olmschenk},
  {Orosz}, {Pal}, {Palle}, {Schwarz}, {Stockdale}, {Tamayo}, {Uhlar}, {Welsh},
  \& {West}}]{Kostov2021}
{Kostov}, V.~B., {Powell}, B.~P., {Torres}, G., {et~al.} 2021, arXiv e-prints,
  arXiv:2105.12586

\bibitem[{{Kratter} {et~al.}(2010){Kratter}, {Matzner}, {Krumholz}, \&
  {Klein}}]{Kratter2010}
{Kratter}, K.~M., {Matzner}, C.~D., {Krumholz}, M.~R., \& {Klein}, R.~I. 2010,
  \apj, 708, 1585

\bibitem[{{Kraus} {et~al.}(2020){Kraus}, {Kreplin}, {Young}, {Bate}, {Monnier},
  {Harries}, {Avenhaus}, {Kluska}, {Laws}, {Rich}, {Willson}, {Aarnio},
  {Adams}, {Andrews}, {Anugu}, {Bae}, {ten Brummelaar}, {Calvet}, {Cur{\'e}},
  {Davies}, {Ennis}, {Espaillat}, {Gardner}, {Hartmann}, {Hinkley}, {Labdon},
  {Lanthermann}, {LeBouquin}, {Schaefer}, {Setterholm}, {Wilner}, \&
  {Zhu}}]{Kraus2020}
{Kraus}, S., {Kreplin}, A., {Young}, A.~K., {et~al.} 2020, Science, 369, 1233

\bibitem[{{Krumholz} \& {McKee}(2020)}]{Krumholz2020}
{Krumholz}, M.~R., \& {McKee}, C.~F. 2020, \mnras, 494, 624

\bibitem[{{Kuffmeier} {et~al.}(2019){Kuffmeier}, {Calcutt}, \&
  {Kristensen}}]{Kuffmeier2019}
{Kuffmeier}, M., {Calcutt}, H., \& {Kristensen}, L.~E. 2019, \aap, 628, A112

\bibitem[{{Lane} {et~al.}(2007){Lane}, {Muterspaugh}, {Fekel}, {Williamson},
  {Browne}, {Konacki}, {Burke}, {Colavita}, {Kulkarni}, \& {Shao}}]{Lane2007}
{Lane}, B.~F., {Muterspaugh}, M.~W., {Fekel}, F.~C., {et~al.} 2007, \apj, 669,
  1209

\bibitem[{{Larson}(1972)}]{Larson1972}
{Larson}, R.~B. 1972, \mnras, 156, 437

\bibitem[{{Larson}(2007)}]{Larson2007}
---. 2007, Reports on Progress in Physics, 70, 337

\bibitem[{{Lee} {et~al.}(2019){Lee}, {Offner}, {Kratter}, {Smullen}, \&
  {Li}}]{Lee2019}
{Lee}, A.~T., {Offner}, S. S.~R., {Kratter}, K.~M., {Smullen}, R.~A., \& {Li},
  P.~S. 2019, \apj, 887, 232

\bibitem[{{Lee} {et~al.}(2020){Lee}, {Offner}, {Hennebelle}, {Andr{\'e}},
  {Zinnecker}, {Ballesteros-Paredes}, {Inutsuka}, \& {Kruijssen}}]{Lee2020}
{Lee}, Y.-N., {Offner}, S. S.~R., {Hennebelle}, P., {et~al.} 2020, \ssr, 216,
  70

\bibitem[{{Manwadkar} {et~al.}(2020){Manwadkar}, {Trani}, \&
  {Leigh}}]{Manwadkar2020}
{Manwadkar}, V., {Trani}, A.~A., \& {Leigh}, N. W.~C. 2020, \mnras, 497, 3694

\bibitem[{{Mardling} \& {Aarseth}(2001)}]{Mardling2001}
{Mardling}, R.~A., \& {Aarseth}, S.~J. 2001, \mnras, 321, 398

\bibitem[{{Mason} {et~al.}(2001){Mason}, {Wycoff}, {Hartkopf}, {Douglass}, \&
  {Worley}}]{WDS}
{Mason}, B.~D., {Wycoff}, G.~L., {Hartkopf}, W.~I., {Douglass}, G.~G., \&
  {Worley}, C.~E. 2001, \aj, 122, 3466

\bibitem[{{Meibom} {et~al.}(2006){Meibom}, {Mathieu}, \&
  {Stassun}}]{Meibom2006}
{Meibom}, S., {Mathieu}, R.~D., \& {Stassun}, K.~G. 2006, \apj, 653, 621

\bibitem[{{M{\'e}nard} {et~al.}(2020){M{\'e}nard}, {Cuello}, {Ginski}, {van der
  Plas}, {Villenave}, {Gonzalez}, {Pinte}, {Benisty}, {Boccaletti}, {Price},
  {Boehler}, {Chripko}, {de Boer}, {Dominik}, {Garufi}, {Gratton}, {Hagelberg},
  {Henning}, {Langlois}, {Maire}, {Pinilla}, {Ruane}, {Schmid}, {van Holstein},
  {Vigan}, {Zurlo}, {Hubin}, {Pavlov}, {Rochat}, {Sauvage}, \&
  {Stadler}}]{Menard2020}
{M{\'e}nard}, F., {Cuello}, N., {Ginski}, C., {et~al.} 2020, \aap, 639, L1

\bibitem[{{Moe} \& {Di Stefano}(2017)}]{Moe2017}
{Moe}, M., \& {Di Stefano}, R. 2017, \apjs, 230, 15

\bibitem[{{Moe} \& {Kratter}(2018)}]{Moe2018}
{Moe}, M., \& {Kratter}, K.~M. 2018, \apj, 854, 44

\bibitem[{{Moe} {et~al.}(2019){Moe}, {Kratter}, \& {Badenes}}]{Moe2019}
{Moe}, M., {Kratter}, K.~M., \& {Badenes}, C. 2019, \apj, 875, 61

\bibitem[{{Naoz}(2016)}]{Naoz2016}
{Naoz}, S. 2016, \araa, 54, 441

\bibitem[{{Offner} {et~al.}(2010){Offner}, {Kratter}, {Matzner}, {Krumholz}, \&
  {Klein}}]{Offner2010}
{Offner}, S. S.~R., {Kratter}, K.~M., {Matzner}, C.~D., {Krumholz}, M.~R., \&
  {Klein}, R.~I. 2010, \apj, 725, 1485

\bibitem[{{Powell} {et~al.}(2021){Powell}, {Kostov}, {Rappaport}, {Borkovits},
  {Zasche}, {Tokovinin}, {Kruse}, {Latham}, {Montet}, {Jensen}, {Jayaraman},
  {Collins}, {Masek}, {Hellier}, {Evans}, {Tan}, {Schlieder}, {Torres},
  {Smale}, {Friedman}, {Barclay}, {Gagliano}, {Quintana}, {Jacobs}, {Gilbert},
  {Kristiansen}, {Colon}, {LaCourse}, {Olmschenk}, {Omohundro}, {Schnittman},
  {Schwengeler}, {Barry}, {Terentev}, {Boyd}, {Schmitt}, {Quinn}, {Vanderburg},
  {Palle}, {Armstrong}, {Ricker}, {Vanderspek}, {Seager}, {Winn}, {Jenkins},
  {Caldwell}, {Wohler}, {Shiao}, {Burke}, {Daylan}, \&
  {Villasenor}}]{Powell2021}
{Powell}, B.~P., {Kostov}, V.~B., {Rappaport}, S.~A., {et~al.} 2021, arXiv
  e-prints, arXiv:2101.03433

\bibitem[{{Pribulla} {et~al.}(2020){Pribulla}, {Puha}, {Borkovits}, {Budaj},
  {Garai}, {Guenther}, {Hamb{\'a}lek}, {Kom{\v{z}}{\'\i}k}, {Kundra},
  {Szab{\'o}}, \& {Va{\v{n}}ko}}]{Pribulla2020}
{Pribulla}, T., {Puha}, E., {Borkovits}, T., {et~al.} 2020, \mnras, 494, 178

\bibitem[{{Quirrenbach} {et~al.}(2019){Quirrenbach}, {Trifonov}, {Lee}, \&
  {Reffert}}]{Quirrenbach2019}
{Quirrenbach}, A., {Trifonov}, T., {Lee}, M.~H., \& {Reffert}, S. 2019, \aap,
  624, A18

\bibitem[{{Raghavan} {et~al.}(2010){Raghavan}, {McAlister}, {Henry}, {Latham},
  {Marcy}, {Mason}, {Gies}, {White}, \& {ten Brummelaar}}]{R10}
{Raghavan}, D., {McAlister}, H.~A., {Henry}, T.~J., {et~al.} 2010, \apjs, 190,
  1

\bibitem[{{Reipurth}(2000)}]{Reipurth2000}
{Reipurth}, B. 2000, \aj, 120, 3177

\bibitem[{{Reipurth} \& {Friberg}(2021)}]{Reipurth2021}
{Reipurth}, B., \& {Friberg}, P. 2021, \mnras, 501, 5938

\bibitem[{{Reipurth} \& {Mikkola}(2012)}]{Reipurth2012}
{Reipurth}, B., \& {Mikkola}, S. 2012, \nat, 492, 221

\bibitem[{{Ricker} {et~al.}(2014){Ricker}, {Winn}, {Vanderspek}, {Latham},
  {Bakos}, {Bean}, {Berta-Thompson}, {Brown}, {Buchhave}, {Butler}, {Butler},
  {Chaplin}, {Charbonneau}, {Christensen-Dalsgaard}, {Clampin}, {Deming},
  {Doty}, {De Lee}, {Dressing}, {Dunham}, {Endl}, {Fressin}, {Ge}, {Henning},
  {Holman}, {Howard}, {Ida}, {Jenkins}, {Jernigan}, {Johnson}, {Kaltenegger},
  {Kawai}, {Kjeldsen}, {Laughlin}, {Levine}, {Lin}, {Lissauer}, {MacQueen},
  {Marcy}, {McCullough}, {Morton}, {Narita}, {Paegert}, {Palle}, {Pepe},
  {Pepper}, {Quirrenbach}, {Rinehart}, {Sasselov}, {Sato}, {Seager},
  {Sozzetti}, {Stassun}, {Sullivan}, {Szentgyorgyi}, {Torres}, {Udry}, \&
  {Villasenor}}]{TESS}
{Ricker}, G.~R., {Winn}, J.~N., {Vanderspek}, R., {et~al.} 2014, in Society of
  Photo-Optical Instrumentation Engineers (SPIE) Conference Series, Vol. 9143,
  Space Telescopes and Instrumentation 2014: Optical, Infrared, and Millimeter
  Wave, ed. J.~{Oschmann}, Jacobus~M., M.~{Clampin}, G.~G. {Fazio}, \& H.~A.
  {MacEwen}, 914320

\bibitem[{{Rohde} {et~al.}(2021){Rohde}, {Walch}, {Clarke}, {Seifried},
  {Whitworth}, \& {Klepitko}}]{Rohde2021}
{Rohde}, P.~F., {Walch}, S., {Clarke}, S.~D., {et~al.} 2021, \mnras, 500, 3594

\bibitem[{{Ryu} {et~al.}(2017){Ryu}, {Leigh}, \& {Perna}}]{Ryu2017}
{Ryu}, T., {Leigh}, N. W.~C., \& {Perna}, R. 2017, \mnras, 467, 4447

\bibitem[{{Sadavoy} \& {Stahler}(2017)}]{Sadavoy2017}
{Sadavoy}, S.~I., \& {Stahler}, S.~W. 2017, \mnras, 469, 3881

\bibitem[{{Sana}(2017)}]{Sana2017}
{Sana}, H. 2017, in The Lives and Death-Throes of Massive Stars, ed. J.~J.
  {Eldridge}, J.~C. {Bray}, L.~A.~S. {McClelland}, \& L.~{Xiao}, Vol. 329,
  110--117

\bibitem[{{Schaefer} {et~al.}(2016){Schaefer}, {Hummel}, {Gies}, {Zavala},
  {Monnier}, {Walter}, {Turner}, {Baron}, {ten Brummelaar}, {Che},
  {Farrington}, {Kraus}, {Sturmann}, \& {Sturmann}}]{Schaefer2016}
{Schaefer}, G.~H., {Hummel}, C.~A., {Gies}, D.~R., {et~al.} 2016, \aj, 152, 213

\bibitem[{{Shatsky}(2001)}]{Shatsky2001}
{Shatsky}, N. 2001, \aap, 380, 238

\bibitem[{{Sterzik} {et~al.}(2003){Sterzik}, {Durisen}, \&
  {Zinnecker}}]{Sterzik2003}
{Sterzik}, M.~F., {Durisen}, R.~H., \& {Zinnecker}, H. 2003, \aap, 411, 91

\bibitem[{{Sterzik} \& {Tokovinin}(2002)}]{Sterzik2002}
{Sterzik}, M.~F., \& {Tokovinin}, A.~A. 2002, \aap, 384, 1030

\bibitem[{{Stone} \& {Leigh}(2019)}]{Stone2019}
{Stone}, N.~C., \& {Leigh}, N. W.~C. 2019, \nat, 576, 406

\bibitem[{{Tobin} {et~al.}(2016){Tobin}, {Kratter}, {Persson}, {Looney},
  {Dunham}, {Segura-Cox}, {Li}, {Chandler}, {Sadavoy}, {Harris}, {Melis}, \&
  {P{\'e}rez}}]{Tobin2016}
{Tobin}, J.~J., {Kratter}, K.~M., {Persson}, M.~V., {et~al.} 2016, Nature, 538,
  483

\bibitem[{{Tokovinin}(2008)}]{Tok2008}
{Tokovinin}, A. 2008, \mnras, 389, 925

\bibitem[{{Tokovinin}(2014)}]{FG67}
---. 2014, \aj, 147, 87

\bibitem[{{Tokovinin}(2016)}]{Tok2016}
---. 2016, \apj, 831, 151

\bibitem[{{Tokovinin}(2017{\natexlab{a}})}]{Tok2017b}
---. 2017{\natexlab{a}}, \mnras, 468, 3461

\bibitem[{{Tokovinin}(2017{\natexlab{b}})}]{Tok2017}
---. 2017{\natexlab{b}}, \apj, 844, 103

\bibitem[{{Tokovinin}(2018{\natexlab{a}})}]{Tok2018b}
---. 2018{\natexlab{a}}, \aj, 155, 160

\bibitem[{{Tokovinin}(2018{\natexlab{b}})}]{MSC}
---. 2018{\natexlab{b}}, \apjs, 235, 6

\bibitem[{{Tokovinin}(2020{\natexlab{a}})}]{Tok2020a}
---. 2020{\natexlab{a}}, \aj, 159, 265

\bibitem[{{Tokovinin}(2020{\natexlab{b}})}]{Tok2020b}
---. 2020{\natexlab{b}}, Astronomy Letters, 46, 612

\bibitem[{{Tokovinin} {et~al.}(2015){Tokovinin}, {Latham}, \&
  {Mason}}]{Tok2015}
{Tokovinin}, A., {Latham}, D.~W., \& {Mason}, B.~D. 2015, \aj, 149, 195

\bibitem[{{Tokovinin} \& {Moe}(2020)}]{TokMoe2020}
{Tokovinin}, A., \& {Moe}, M. 2020, \mnras, 491, 5158

\bibitem[{{Tokovinin} {et~al.}(2006){Tokovinin}, {Thomas}, {Sterzik}, \&
  {Udry}}]{Tok2006}
{Tokovinin}, A., {Thomas}, S., {Sterzik}, M., \& {Udry}, S. 2006, \aap, 450,
  681

\bibitem[{{Tokovinin}(1997)}]{MSC1}
{Tokovinin}, A.~A. 1997, \aaps, 124, 75

\bibitem[{{Tokovinin} \& {Smekhov}(2002)}]{TS2002}
{Tokovinin}, A.~A., \& {Smekhov}, M.~G. 2002, \aap, 382, 118

\bibitem[{{Toonen} {et~al.}(2020){Toonen}, {Portegies Zwart}, {Hamers}, \&
  {Bandopadhyay}}]{Toonen2020}
{Toonen}, S., {Portegies Zwart}, S., {Hamers}, A.~S., \& {Bandopadhyay}, D.
  2020, \aap, 640, A16

\bibitem[{{Tremaine}(2020)}]{Tremaine2020}
{Tremaine}, S. 2020, \mnras, 493, 5583

\bibitem[{{Valtonen} \& {Karttunen}(2006)}]{Valtonen2006}
{Valtonen}, M., \& {Karttunen}, H. 2006, {The Three-Body Problem} (Cambridge
  University Press)

\bibitem[{{V{\'a}zquez-Semadeni} {et~al.}(2019){V{\'a}zquez-Semadeni}, {Palau},
  {Ballesteros-Paredes}, {G{\'o}mez}, \& {Zamora-Avil{\'e}s}}]{Vasquez2019}
{V{\'a}zquez-Semadeni}, E., {Palau}, A., {Ballesteros-Paredes}, J.,
  {G{\'o}mez}, G.~C., \& {Zamora-Avil{\'e}s}, M. 2019, \mnras, 490, 3061

\bibitem[{{Whitworth}(2001)}]{Whitworth2001}
{Whitworth}, A.~P. 2001, in The Formation of Binary Stars, ed. H.~{Zinnecker}
  \& R.~{Mathieu}, Vol. 200, 33

\bibitem[{{Wolf} {et~al.}(2018){Wolf}, {Ku{\v{c}}{\'a}kov{\'a}}, {Zasche},
  {Vra{\v{s}}til}, {Ho{\v{n}}kov{\'a}}, {Hornoch}, {Lehk{\'y}}, {Ma{\v{s}}ek},
  {{\v{S}}melcer}, {Tyl{\v{s}}ar}, {Nov{\'a}k}, {{\v{C}}ervinka}, \&
  {B{\v{e}}l{\'\i}k}}]{Wolf2018}
{Wolf}, M., {Ku{\v{c}}{\'a}kov{\'a}}, H., {Zasche}, P., {et~al.} 2018, \aap,
  620, A72

\bibitem[{{Worley}(1967)}]{Worley1967}
{Worley}, C.~E. 1967, in On the Evolution of Double Stars, ed. J.~{Dommanget},
  Vol.~17, 221

\bibitem[{{Xia} {et~al.}(2019){Xia}, {Fu}, \& {Wang}}]{Xia2019}
{Xia}, F., {Fu}, Y., \& {Wang}, X. 2019, \apj, 882, 147

\bibitem[{{Yuan} {et~al.}(2020){Yuan}, {Li}, {Zhu}, {Liu}, {Wang}, {Liu},
  {Kim}, {Tatematsu}, {Yuan}, \& {Wu}}]{Yuan2020}
{Yuan}, L., {Li}, G.-X., {Zhu}, M., {et~al.} 2020, \aap, 637, A67

\bibitem[{{Zasche} {et~al.}(2019){Zasche}, {Vokrouhlick{\'y}}, {Wolf},
  {Ku{\v{c}}{\'a}kov{\'a}}, {K{\'a}ra}, {Uhla{\v{r}}}, {Ma{\v{s}}ek}, {Henzl},
  \& {Caga{\v{s}}}}]{Zasche2019}
{Zasche}, P., {Vokrouhlick{\'y}}, D., {Wolf}, M., {et~al.} 2019, \aap, 630,
  A128

\bibitem[{{Z{\'u}{\~n}iga-Fern{\'a}ndez}
  {et~al.}(2021){Z{\'u}{\~n}iga-Fern{\'a}ndez}, {Olofsson}, {Bayo}, {Haubois},
  {Corral-Santana}, {Lopera-Mej{\'\i}a}, {Ronco}, {Tokovinin}, {Gallenne},
  {Kennedy}, \& {Berger}}]{ZF2021}
{Z{\'u}{\~n}iga-Fern{\'a}ndez}, S., {Olofsson}, J., {Bayo}, A., {et~al.} 2021,
  arXiv e-prints, arXiv:2109.02841

\end{thebibliography}

\end{document}